\documentclass[a4paper,11pt]{article}
\usepackage{jinstpub} 
\usepackage{lineno}
\usepackage{booktabs} 
\usepackage{xspace} 
\usepackage{wasysym} 

\title{\boldmath 
    Optimization of 3D diamond detectors with graphitized electrodes based on an innovative numerical simulation.
    }

\usepackage{xcolor}
\newcommand\qsme{Maxwell equations in the quasi-static limit\xspace}

\newcommand\comsol{COMSOL MultiPhysics\textsuperscript{\tiny\textregistered}\xspace}
\newcommand\ohmcm{\ensuremath{\Omega\cdot\mathrm{cm}}\xspace}
\newcommand\ii{\mathrm{i}} 
\newcommand\methodA{\ensuremath{\mathrm{Method\, A}}\xspace}
\newcommand\methodB{\ensuremath{\mathrm{Method\, B}}\xspace}
\newcommand\methodC{\ensuremath{\mathrm{Method\, C}}\xspace}

\author[a]{L.~Anderlini}
\author[a]{A.~Bombini}
\author[*,a,b]{C.~Buti}
\author[c]{D.~Janssens}
\author[b]{S.~Lagomarsino}
\author[a]{G.~Passaleva}
\author[a,d]{M.~Veltri}

\affiliation[a]{INFN Firenze,\\ via G. Sansone 1, Sesto Fiorentino, Italy}
\affiliation[b]{University of Florence, Physics and Astronomy Department,\\ via G. Sansone 1, Sesto Fiorentino, Italy}
\affiliation[c]{European Council for Nuclear Research,\\ Espl. des Particules 1, Genève, Switzerland}
\affiliation[d]{University of Urbino, Pure and Applied Science Department,\\ via Santa Chiara 27, Urbino, Italy}
\emailAdd{Clarissa.Buti@fi.infn.it}

\abstract{ 
Future experiments at hadron colliders require an evolution of the tracking sensors to ensure sufficient radiation hardness as well as space and time resolution to handle unprecedented particle fluxes. 3D diamond sensors with laser-graphitized electrodes are promising candidates due to their strong binding energy, small atomic number, and high carrier mobility. However, the high resistance of the engraved electrodes delays the propagation of the induced signals towards the readout electronics, thereby degrading the precision of the timing measurements. So far, this effect has been the dominant factor limiting the time resolution of these devices, with other contributions, such as those due to electric field inhomogeneities or electronic noise, typically neglected. Recent advancements in graphitization technology, however, motivate a renewed effort in modeling signal generation in 3D diamond detectors, to achieve more reliable predictions. To this purpose, we apply an extended version of the Ramo-Shockley theorem, describing the effect of signal propagation as a time-dependent weighting potential, obtained by numerically solving the Maxwell’s equations in a quasi-static approximation. We developed a custom spectral method solver and validated it against \texttt{\comsol}. The response of the modeled sensor to a beam of particles is then simulated using \texttt{Garfield++} and is compared to the data acquired in a beam test carried on in 2021 by the TimeSPOT Collaboration at the SPS, at CERN. Based on the results obtained with this simulation workflow, we conclude that reducing the resistivity of the graphitic columns remains the priority for significantly improving the time resolution of 3D diamond detectors. Once achieved, optimization of the detector geometry and readout electronics design will become equally important steps to further enhance the timing performance of these devices.
}


\begin{document}
\maketitle
\flushbottom

\section{Introduction}
\label{sec:introduction}

One of the key challenges in the design of detectors for future particle physics experiments is ensuring their reliable operation under extreme particle flux conditions. 
This is required by the need of operating experiments at very high instantaneous luminosity to collect high statistics physics event samples. 

The development of the ATLAS and CMS detectors for the High-Luminosity 
Large Hadron Collider~\cite{DaVia:2012ay, CMSTracker:2019aux, DallaBetta:2016czx, DallaBetta:2016dqn},
the upcoming upgrade of the LHCb experiment~\cite{LHCB-TDR-023, LHCB-TDR-026}, and the detectors at 
the hadronic version of the Future Circular Collider (FCC-hh)~\cite{FCC:2018vvp}, or at the Muon 
Collider~\cite{InternationalMuonCollider:2024jyv} exemplify the growing demand for sensors 
capable of withstanding an unprecedented flux of ionizing particles. 

Similar challenges arise in technological and applied research, such as in the development of dosimeters 
for Flash Radiotherapy~\cite{Porter:2023crm,frontiers-review} or in providing precise process monitors  
in nuclear fission and fusion facilities~\cite{Metcalfe:2017slh}.

Physics requirements for experiments at future high luminosity colliders impose severe constraints especially on tracking detectors placed close to the interaction points. Besides very high radiation hardness, they must feature pixels with excellent time resolution to ensure an efficient tracking and micrometer-level precision in primary and secondary vertex reconstruction~\cite{LHCB-TDR-026}.

Among the many technologies that are being considered, like for example the recently proposed doping compensation in LGAD silicon sensors~\cite{Sola:2022dyc}, diamond detectors have been extensively studied as highly promising candidates 
for addressing the challenges posed by extreme operating conditions.
Diamond sensors offer indeed several advantages. Firstly, they have enhanced radiation tolerance due to the strong lattice binding energy and relatively low atomic number. Secondly,
the high mobility and saturation velocity of charge carriers provide faster 
signal collection, leading to improved time resolution compared to silicon devices~\cite{Pomorski2009,pernegger}. 
Moreover, diamond sensors have a sufficiently large band-gap to make them solar-blind and operable at higher 
temperatures. This characteristic is particularly advantageous for applications where cooling systems 
are impractical due to space or complexity constraints.

An intriguing feature of diamond is its ability to undergo laser-induced graphitization, 
transforming few $\mathrm{\mu m^3}$ of diamond around the laser focus into a conductive 
admixture of graphite and amorphous carbon. 
This process enables the fabrication of full-carbon sensors with high flexibility in the geometry of electrodes which can be realized either on the surface or in the bulk of the diamond crystals~\cite{Oliva:2019alx,PORTER2023109692,Anderlini:2021pei,Bachmair:2015iba}.

\begin{figure}
    \begin{minipage}[t]{0.45\textwidth}
        \centering
        \includegraphics[width=\textwidth]{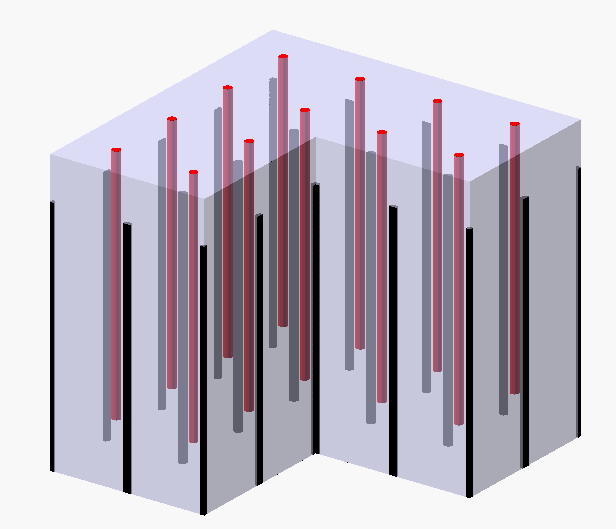}
        \caption{\label{fig:schematic-view}
            Diagrammatic illustration of a three-dimensional diamond sensor, depicting a segment comprising 
            four by four fundamental units. 
            The figure illustrates electrodes connected to the polarization voltage (black) 
            and grounded terminals (red).
            Proportions are illustrative; actual dimensions may vary.
        }
    \end{minipage}
    \hfill
    \begin{minipage}[t]{0.45\textwidth}
        \centering
        \includegraphics[width=\textwidth]{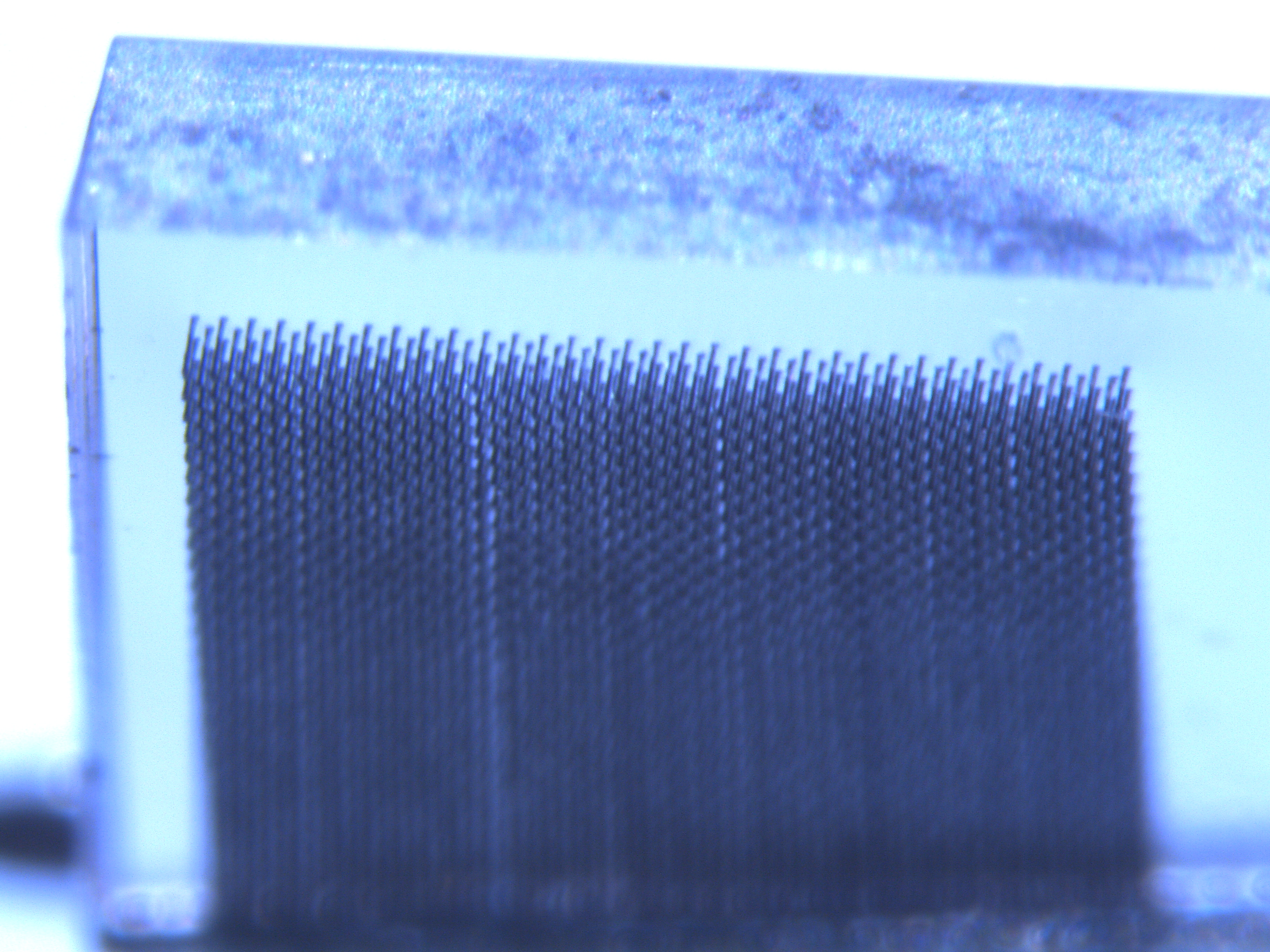}
        \caption{\label{fig:photo-sensor}
            A detailed microscopic image of a 3D diamond sensor is presented. 
            The specimen was slightly tilted during acquisition to ensure the complete 
            visualization of the graphitized electrodes.
        }
    \end{minipage}
\end{figure}

Similarly to the silicon sensors, the 3D geometry, where electrodes are perpendicular to the sensor surface, offers significant advantages over the traditional planar geometry. In this configuration, the drift length of charge carriers is substantially shorter with respect to the planar geometry, reducing the probability of charge trapping by radiation induced defects, and shortening the charge collection time, leading to faster signals. A sketch of a 3D electrode arrangement and a picture of a 3D diamond sensor are shown in Figure~\ref{fig:schematic-view} and Figure~\ref{fig:photo-sensor} respectively.

The fabrication of 3D diamond sensors presents substantial challenges due to the high refractive 
index of diamond, which degrades the laser focus due to spherical aberration, and because of the major resistance of the diamond bulk
to laser-induced phase transitions. 
The transition from diamond to graphite within the bulk is complicated by differences in density between 
the two materials, leading to mechanical strain in the crystal lattice. 
Despite these challenges, significant progress has been made. For instance, Salter \emph{et al.} 
demonstrated in 2017 that phase transitions in the diamond bulk occur through discrete nano-scale 
structural modifications aligned with the diagonals of the diamond lattice~\cite{salter2017high}. 
The electrical resistivity of these graphitized regions is influenced by random connections 
between conductive structures and is typically orders of magnitude higher than that of 
graphitized surfaces~\cite{Kononenko_2015}.

Several other studies, across various scientific communities, proposed methods to reduce the resistivity of the engraved 
structures. 
The most promising results have been achieved using femtosecond infrared lasers, where spherical aberration
introduced by the diamond surface is corrected through Spatial Light 
Modulators~\cite{Anderlini:2022exf,test-beam-sensors}. 
This approach has enabled the fabrication of structures parallel to the diamond surface with a recorded 
resistivity of 0.02~\ohmcm\cite{Salter:2014ohm}. Additionally, the same experimental setup has 
demonstrated the capability to engrave multiple electrodes simultaneously~\cite{Krueger:24}.
On the other hand, employing Bessel femtosecond-pulsed beams without adaptive 
spherical aberration correction has resulted in the formation of fine electrodes orthogonal to the 
diamond surface, exhibiting a resistivity of 0.04~\ohmcm~\cite{KURIAKOSE2023110036}.

High resistivity in graphitized structures leads to considerable electrode resistance in diamond sensors. 
The increased resistance affects the signal propagation time through the electrodes. While the dependence of the propagation delay on the depth of the deposited charge can usually be neglected in the case of electrodes that can be assumed to be perfectly conductive (\emph{metallic electrodes} hereafter), it becomes the dominant factor limiting the time resolution in 3D diamond sensors. Until recently, such a contribution to the time resolution from signal propagation was so large compared to those from carrier transport and electronic noise that could be effectively considered as the sole contribution, allowing for the use of simplified models.
However, 3D laser-graphitized diamond sensors with a time resolution significantly below 100 ps have been reported~\cite{Anderlini:2022exf}, thanks to the advanced graphitization techniques mentioned above, strongly motivating a 
renewed effort to model the signal formation in diamond detectors by properly taking into account the combined effect of carrier transport through the bulk and signal propagation through resistive electrodes.

In this work, we propose an innovative numerical model based on an extension of the Ramo-Shockley theorem to conductive media to simulate the signal induction process in 3D diamond sensors with graphitic electrodes, including the contribution of the resistive elements as described in Refs.~\cite{Gatti1982, Riegler2002, Rigler:2e004jh,Riegler2019,Janssens:2890572}.

To validate our proposed method, we use experimental data obtained from a beam test conducted at CERN, 
thereby ensuring the reliability and applicability of our model. 
Having validated the method, we simulated sensors with different geometries and configurations to study their time resolution.  The simulations allowed us to disentangle the various contributions to the time resolution, enabling a systematic investigation aimed at identifying potential optimizations.

\section{Signal simulation in 3D diamond sensors with graphitic electrodes}
\label{sec:sim}

Particles traversing a radiation detector release part of their energy inside the detection medium, generating charge carriers which drift towards the electrodes, inducing a current signal. 

The simulation of signal generation involves several processes, which will be discussed in detail throughout this section:  

\begin{enumerate}
    \item Interaction of an ionizing particle with the detector medium;
    \item Transport of the generated charge carriers through the medium;
    \item Propagation along the electrodes of the signal resulting from charge carrier motion; 
    \item Amplification and shaping introduced by the readout front-end electronics.
\end{enumerate}

This work does not introduce novelties in the simulation of the processes 1, 2 and 4, which are briefly reviewed in the following to provide a complete overview of simulation procedure and to report on the technical choices adopted in the deployment of the simulation.
Process 3 is rarely relevant for sensors with metallic electrodes because the propagation of the signal through the perfectly conductive material is instantaneous compared to the time scale of process 2.

The simulation of process 1 and process 2 is presented in Section~\ref{sec:transport}, while Section~\ref{sec:metallicelectrodes} reviews the description of the signal induction in the simplified case of metallic electrodes, by leveraging the Ramo-Shockley theorem.

As mentioned in the introduction, however, the high resistivity of graphitized electrodes makes the contribution to the time resolution from process 3 to dominate in the case of 3D diamond sensors.

To model the effect of process 3 on the induced signal we discuss two approaches:
\begin{itemize}
    \item approximating the effect of resistive electrodes with a discrete impedance network, following the equivalent circuit principle, hereafter referred to as \methodA (Section~\ref{sec:cellmodel});
    \item using a time-dependent extension of the Ramo-Shockley theorem (Section~\ref{sec:ramo}), in which two different methods to compute the electric field maps have been developed: a consolidated one based on the Finite Element Method (FEM, \methodB), and an innovative one based on a pseudo-spectral method for solving the differential equations (\methodC).
\end{itemize}

In subsection~\ref{sec:comparison}, a comparison between all the simulation methods is provided, along with an explanation of the final steps related to noise and the simulation of the front-end electronics, required to process the output signals from all simulations so that they can be compared with experimental data for validation purposes.  

Finally, to enable comparison of the simulation with experimental data, we discuss the simulation of the effects introduced by the electronics (process 4) in Section~\ref{sec:kuboard}.

The simulation of processes 1 and 2 relies on the {\texttt{Garfield++}} software package~\cite{Schindler:2012wta}. The simulation of process 3 with \methodB and \methodC have also been developed within the \texttt{Garfield++} framework and were made available in the official repository of the project.

The sensor considered for the simulation is the 3D diamond sensor described in Ref.~\cite{Anderlini:2022exf}, engraved in a 500~$\mathrm{\mu m}$ thick diamond specimen and featuring 12~$\mathrm{\mu m}$-diameter graphitized electrodes, organized on a two-dimensional $55\times55$~$\mathrm{\mu m^2}$-pitch pixel matrix. 
The readout electrodes are arranged in square cells, with polarization electrodes placed at the four corners and the readout electrode at the center. 
Both polarization and readout electrodes are 450~$\mathrm{\mu m}$ long and are engraved starting from opposite sides of the crystal. 
Their resistance was indirectly measured to be 30~k$\mathrm{\Omega}$, corresponding to a resistivity of 0.75~$\mathrm{\Omega\cdot cm}$. 

The sensor was selected for benchmarking the simulation based on a campaign of measurements conducted in 2022 using a 180 GeV pion beam, which provided experimental data for comparison with simulation results~\cite{Anderlini:2022exf}.

\subsection{Computing trajectories of charge carriers with \texttt{Garfield++}}
\label{sec:transport}
The numerical model proposed in this work is built upon established methodologies 
for simulating the migration of charge carriers generated by the passage of ionizing 
particles. The motion of charges occurs within the electrostatic field $\vec E$ defined by the 
polarization voltage $V_p$ applied to sensor electrodes.

The ionization charge generated by the traversing particle is simulated using the High Energy Electrodynamics (\texttt{Heed}) toolkit \cite{Smirnov2005,Heed} as integrated in \texttt{Garfield++}. 
The detector response was investigated through simulations of tracks produced by 180~GeV/$c$ pions impinging on the sensor surface and producing charge clouds of approximately 15000 electrons distributed in clusters along the particle path. The stochastic nature of the energy deposition is intrinsically taken into account in \texttt{Heed}, 
where the local fluctuations of the ionization charge along the particle track follow a Landau-like distribution~\cite{Landau:1944if}. 
This approach ensures a realistic representation of the spatial non-uniformity of the charge generation process within the sensor volume.

The drift of generated electrons and holes is determined by the electrostatic field, which determines both their trajectories toward the electrodes and their velocities, calculated by accounting for the carrier mobility ($\mu$) and saturation velocity ($v_s$) in diamond:

\begin{equation}
    \label{eq:velocity}
    \vec{v}(t) = \frac{\mu \vec{E}}{1+\frac{\mu E}{v_s}} .
\end{equation}

Due to the complex structure of the 
graphitized electrodes, carrier drift
is neglected within the regions representing the conductive carbon phase. 
It should be noted that this simplification introduces discrepancies when the ionizing particle 
traverses an electrode because the deposited charge is not accounted for in our simulation model. 
However, beam test results on 3D diamond sensors show tracking efficiencies above 99\%, 
indicating that the charge deposited in the
graphitized material effectively contributes 
to the signals collected at the electrode ends~\cite{Anderlini:2022exf}.

Different methods were used to compute the map of the electrostatic potential (V) by solving the Partial Differential 
Equation (PDE),
\begin{equation}
    \label{eq:Poisson}
    \nabla^2 V(\vec x) = -\frac{\rho(\vec x)}{\epsilon_0\epsilon_r}
    \quad\mathrm{with}\quad
    \left\{
        \begin{array}{ll}
            V(\vec x) = V_p & \mbox{for $\vec x$ within the polarization electrodes} \\
            V(\vec x) = 0 & \mbox{for $\vec x$ within the read-out electrodes} \\
            \rho(\vec x) = 0 & \mbox{for $\vec x$ within the dielectric volume} \\
        \end{array}
    \right.,
\end{equation}
leading to consistent carrier trajectories once deployed in \texttt{Garfield++}. Here $\epsilon_0$ and $\epsilon_r$
are respectively the vacuum and the relative permittivity, and $\rho(\vec{x})$ is the charge density at point $\vec{x}$.

The results described below were obtained computing the electrostatic potential in the diamond sensor with either 
a FEM implemented in
\texttt{\comsol}~\cite{comsol} on a custom mesh (\methodB)~\cite{Janssens:2890572},
or using an iterative spectral method on a uniform voxel grid, resetting
the boundary conditions on voltage and charge density at each iteration (\methodC).

\begin{figure}
    \centering
    \includegraphics[width=0.7\textwidth,clip,trim=10mm 30mm 10mm 30mm]{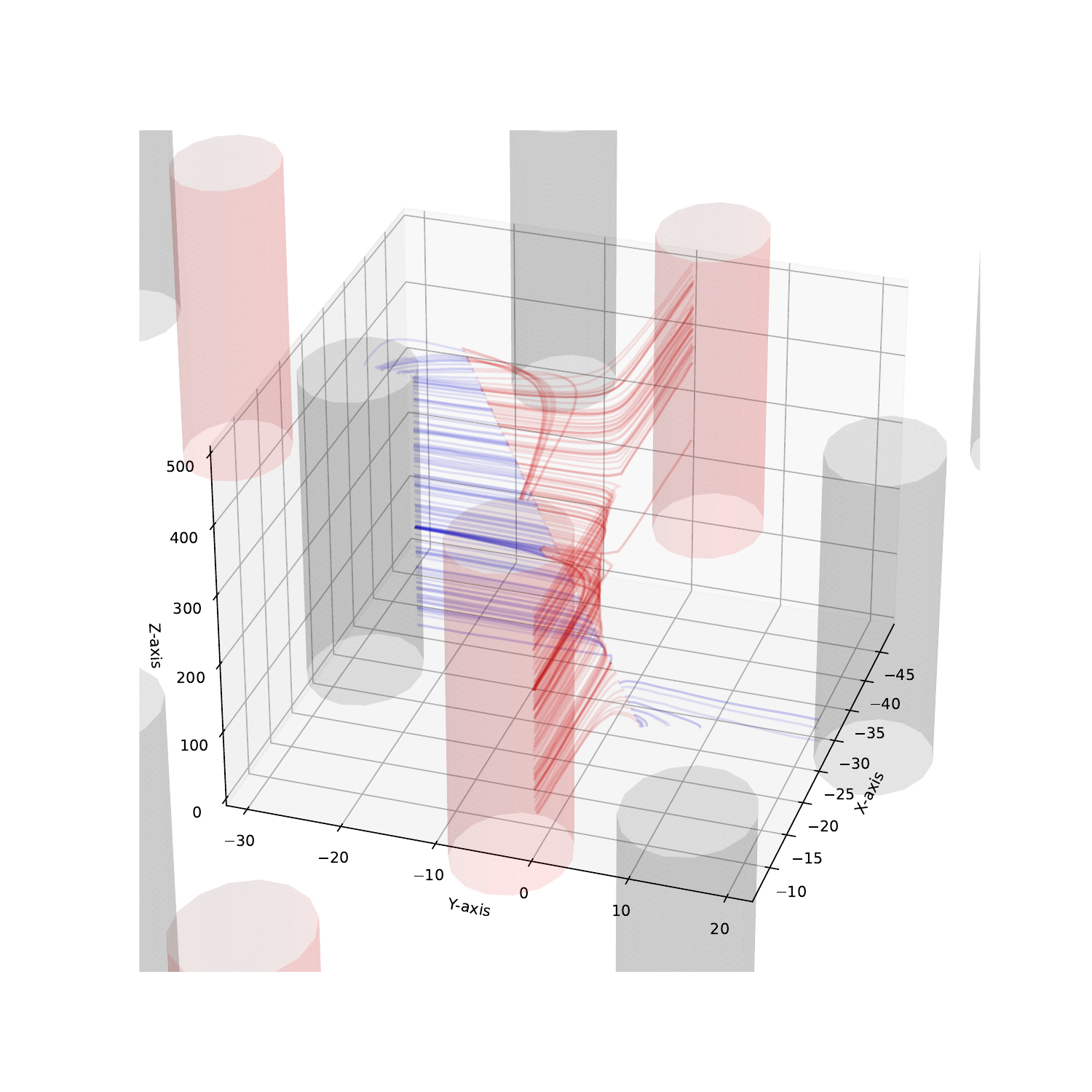}
    \caption{
        \label{fig:paths}
        Simulated trajectories of electrons (blue lines) and holes (red lines) generated 
        by a 180~GeV/$c$ pion traversing a diamond sensor. 
        The graphitized electrodes are depicted as gray or red cylinders, representing their connection to 
        polarization voltage (positive in this case) and readout (grounded), respectively. 
        The electrostatic field was derived by solving the Poisson equation using \texttt{\comsol}, 
        with drift paths computed by \texttt{Garfield++}.
    }
\end{figure}

The paths of the carriers deposited by a typical traversing particle in a 3D diamond device are
shown in Figure~\ref{fig:paths}.

\subsection{Signal induced on ideal, perfectly conductive electrodes}
\label{sec:metallicelectrodes}
According to the Ramo-Shockley theorem \cite{Ramo1939,Shockley1938}, the current induced on a given electrode by the drift of a charge $q$ along a path $\vec{x}(t)$ with velocity $\vec{v}(t)$ is given by
\begin{equation}
    \label{eq:ramoshockley}
    i(t) = -\frac{q}{V_w} \vec{E}_w(\vec{x}) \cdot \vec{v}(t).
\end{equation}
Here, $\vec{E}_w$ denotes the weighting field, obtained by solving Laplace’s equation for the sensor geometry with a fixed potential $V_w$ applied to the electrode of interest, while all other electrodes are set to ground. Under static conditions, the electric potential is assumed to be uniform within each electrode, allowing the Dirichlet problem to be solved using the potential at the interface between the electrodes and the diamond as boundary conditions.

Neglecting the impedance of the electrodes, the weighting field $\vec{E}_w$ can be considered as time-independent. In order to mark a distinction from the time-dependent weighting potential that will be discussed later, it is customary to refer to $\vec{E}$ as the \emph{steady weighting field}. 

According to the prescription of the Ramo–Shockley theorem, this steady weighting field was evaluated using both numerical methods B and C. The resulting signal corresponds to the case in which the sensor electrodes can be considered metallic.

\begin{figure}
    \centering
    \includegraphics[width=0.7\textwidth]{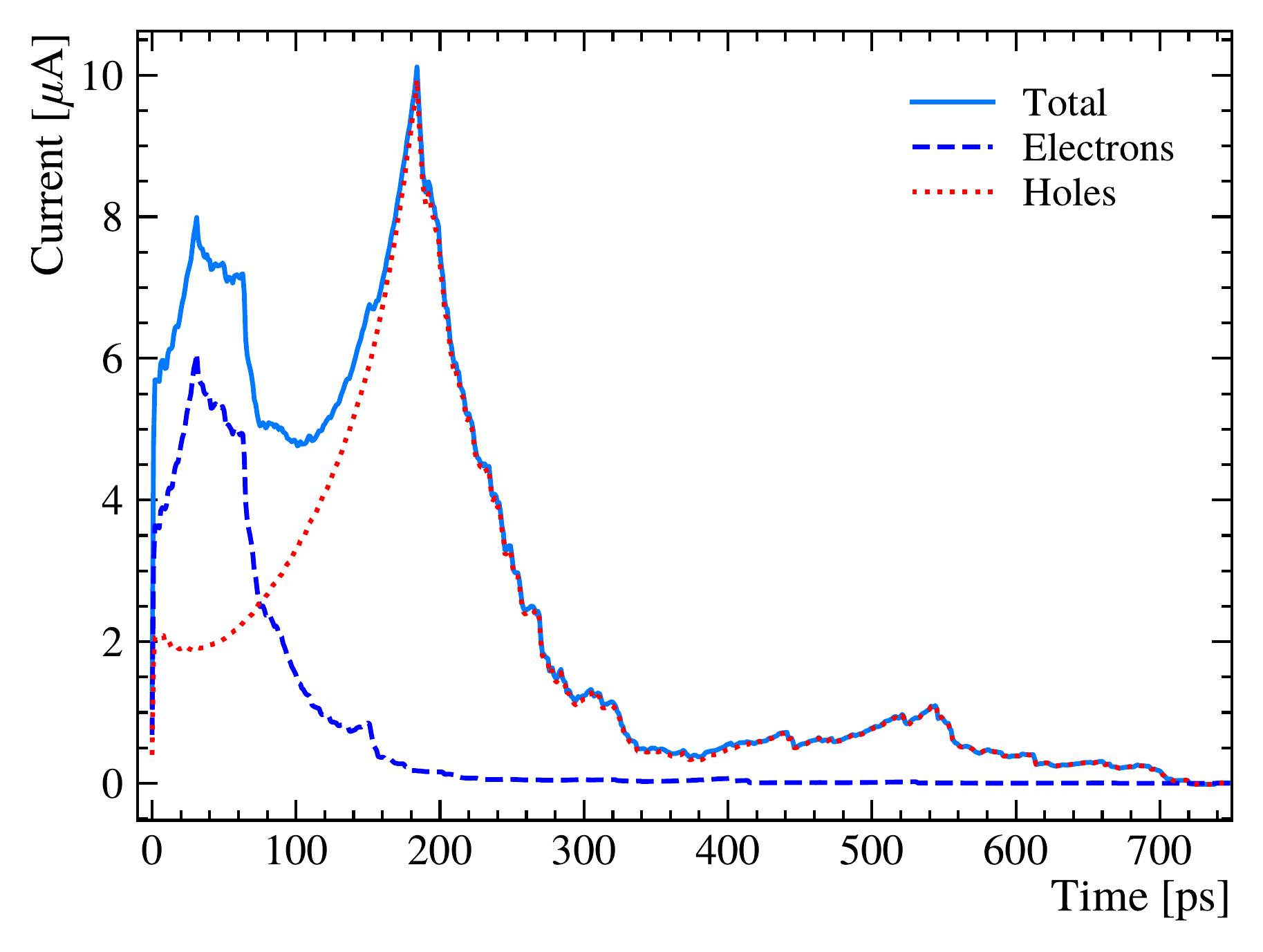}
    \caption{\label{fig:signal-steady}
        Signal induced on perfectly conductive electrodes embedded in a diamond 
        dielectric traversed by a 180~GeV/$c$ pion.
        The signal model includes only processes 1 and 2, excluding propagation effects (process 3).
        The contributions due to the induction of electrons and holes are 
        presented separately.
        }
\end{figure}

An example of a signal obtained with this method for the 3D geometry described at the beginning of 
Section~\ref{sec:sim} is shown in Figure~\ref{fig:signal-steady}, where the electron and hole contributions are displayed separately. The signal spans approximately 300 picoseconds and exhibits two distinct peaks corresponding to the motion of the electron and hole charge distributions in regions of high weighting field, close to their respective collection electrodes. The shape of these signals depends in general on the trajectory of the ionizing particle with respect to the readout electrodes.

\subsection{Modeling resistive electrodes with a discrete impedance network (\methodA)}\label{sec:cellmodel}
As a preliminary step, in order to have a simplified benchmark available to gauge our results with, the system of a readout and a polarization graphitic electrode is described as a discrete impedance network, where the current signal induced by the drift of charge carriers inside the sensor is injected and propagated to the readout electronics input.

\begin{figure}
    \centering
    \includegraphics[width=\textwidth]{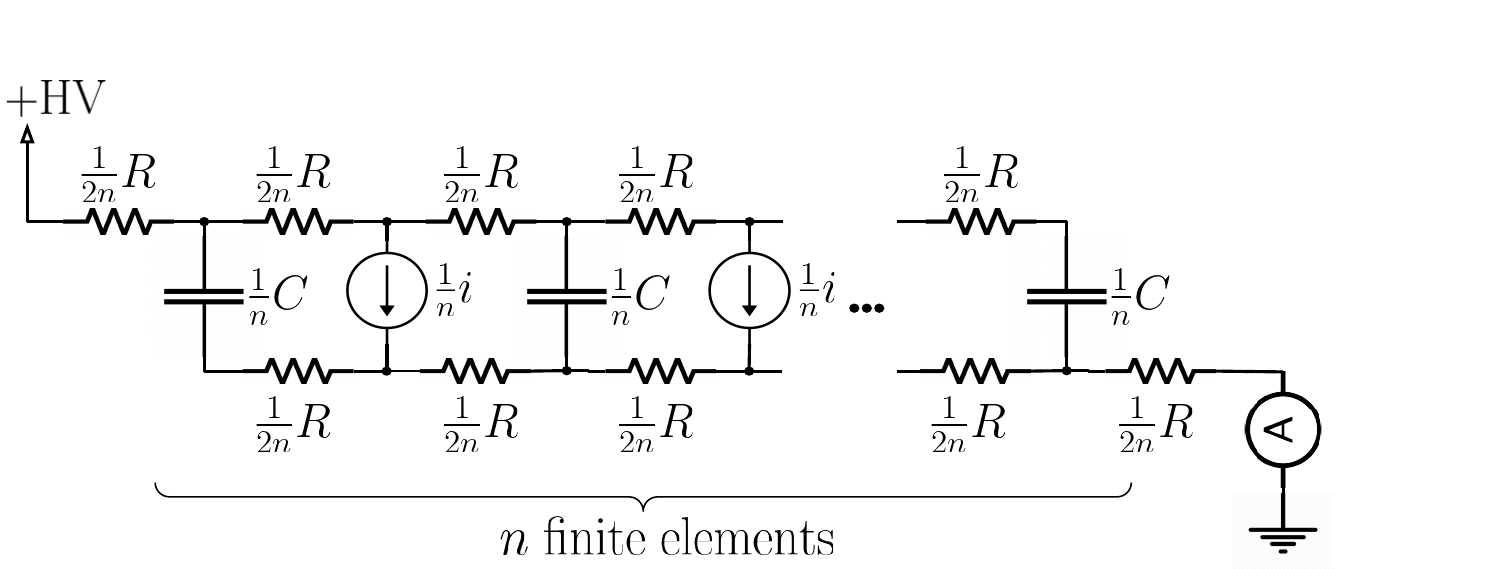}
    \caption{\label{fig:cell-model-schematics}
        A schematic representation of two electrodes used for detecting a traversing particle, depicted as an 
        impedance network. 
        Ionization charges are modeled using idealized pulse current generators. 
        The total capacitance ($C$) and resistance ($R$) are distributed uniformly across $n$
        identical elements.
    }
\end{figure}

A crude approximation to take into account the effect of signal propagation through the graphitized electrodes is to model them as a discrete ladder of capacitors and resistors, as shown in Figure~\ref{fig:cell-model-schematics}. 

The value of the resistance of each element can be computed by distributing uniformly
through the elements the full resistance of the graphitized column.
Experimental measurements indicate a value of the resistance of each column of the 
order of $10^4 - 10^5~\Omega$.

The capacitance of an elementary cell of the sensor depends on the column diameter, 
which is difficult to assess with precision because of the nano-structured nature of the graphitized material. 
Experimental measurements on fabricated sensors are also subject to large cell-to-cell variations. 
Therefore, the order of magnitude of the capacitance, as computed with numerical models and confirmed 
by measurements, is in the range of $10-100$~fF.

From the equivalent circuit shown in Figure~\ref{fig:cell-model-schematics}, 
a delta response function — i.e. the transfer function of the discrete-impedance network — is obtained by applying an elementary pulse in the current generators and measuring 
the current flowing towards ground. 
To account for the fluctuations of the ionization charge deposited through the depth of the columns (process 1), 
the elementary current pulse is split randomly through the discrete elements of the model according 
to the Landau distribution~\cite{Landau:1944if}. 
The transfer function is computed using the 
\texttt{Ngspice} software package~\cite{ngspice} and is then 
convolved numerically with the signal obtained with \texttt{Garfield++} using the steady
weighting field.

The discrete impedance network model provides a fast, heuristic but fairly reasonable estimate of the effect of the electrode resistivity on the signals~\cite{Anderlini:2021pei, Anderlini:2022exf}. However, it is limited to symmetric geometries and relies on idealized assumptions.
First of all one has to consider the large uncertainties on the capacitance and the approximations coming in particular  from neglecting the correlation effects between charge-carrier trajectories and signal propagation. In addition, the model is valid only insofar as the polarization
and readout electrodes can be considered symmetric and that the shaping time constant of the readout electronics is of the same order of magnitude of the charge-carrier transport timescale, or larger.

In the following, we present a novel approach that, while computationally intensive and technically demanding, offers a more complete and theoretically robust insight into the physical processes underlying signal formation in 3D diamond sensors. Comparing predictions from this advanced method with those from the simplified model helps deepen the understanding of sensor behavior, especially in regimes where physical intuition still provides meaningful guidance.

\subsection{Time-dependent extension of the Ramo-Shockley theorem}\label{sec:ramo}

The propagation of the signal through the resistive electrodes can be rigorously described by using an extended form of the Ramo-Shockley theorem for conductive media. The contribution of the material resistivity to the signal formation is included in the time evolution of the weighting potential, which is the solution of the Maxwell equations in the quasi-static limit~\cite{Rigler:2e004jh}.
\begin{equation}
    \label{eq:quasistaticmaxwell-system}
    \left\{
        \begin{array}{l}
            \nabla^2 \Psi(\vec x, t) = -\frac{\rho (\vec x, t)}{\epsilon}\\
            \vec \nabla \cdot (\sigma(\vec x)\, \vec \nabla \Psi(\vec x, t)) - \frac{\partial \rho(\vec x, t)}{\partial t} = 0
        \end{array}
    \right.
\end{equation}
where $\Psi$ represents the weighting potential, such that $\vec{E}_w = - \nabla\Psi$, $\sigma$ the electrical conductivity, 
$\epsilon$ the dielectric constant and $\rho$ the charge distribution.
The latter is not of immediate interest in the computation of the time-dependent 
weighting potential. It can be treated implicitly by combining the Poisson and continuity 
equations into a single equation:
\begin{equation}
    \label{eq:quasistaticmaxwell-equation}
    \frac{\partial}{\partial t} \nabla^2 \Psi(\vec x, t) + \frac{1}{\epsilon} \vec \nabla \cdot (\sigma(\vec x)\, \vec \nabla \Psi(\vec x, t)) = 0.
\end{equation}

The current induced by an elementary charge $q$, that is expressed by Eq.~\ref{eq:ramoshockley} 
for perfectly conductive electrodes, is now obtained through a convolution of the transport (process 2) and propagation effects (process 3)
\begin{equation}
    \label{eq:ramoshockleyriegler}
    i(t) = -\frac{q}{V_w}\int_0^t \vec H(\vec x(t'), t-t') \cdot \vec v(t') \mathrm{d} t'
    \qquad \mathrm{with} \qquad
   \vec{H}(\vec{x}, t) = -\nabla \frac{\partial \Psi(\vec{x}, t) \Theta(t)}{\partial t} \, .
\end{equation}
where $\vec{v}(t)$ is always the one defined in Equation~\ref{eq:velocity}.

With this approach, the simulation of the transport of the charge carriers can be decoupled
from the simulation of the propagation effects that are condensed in time-dependent weighting 
potential maps that only depend on the sensor geometry, encoded in the conductivity 
$\sigma(\vec x)$ and dielectric constant $\epsilon$.
It is worth noticing that Equation~\ref{eq:quasistaticmaxwell-equation} is invariant for 
transformations $t \to t/\alpha$ and $\sigma(\vec x) \to \alpha \sigma(\vec x)$ $\forall \alpha \in \mathbb R^+$,
which implies that a reduction of resistivity (\emph{i.e}., an increase in conductivity) results into a linear reduction of the timescale of the signal propagation, for any fixed geometry. 
In addition, since $\Psi(\vec x, t/\alpha; \alpha\sigma(\vec x)) = \Psi(\vec x, t; \sigma(\vec x))$, the numerical solutions of the equation can be reinterpreted by scaling the time 
axis instead of solving the equation again for a rescaled resistivity. 

Calculating the time-dependent weighting potential via the \qsme constitutes 
a well-defined problem. 
Although computationally intensive because of the third-order PDE and of the
complicated geometry, numerical solutions of Cauchy problems benefit from extensive research 
and development across various sectors, offering numerous methodologies. 
In the subsequent sections, we present a comparative analysis of results derived from 
\texttt{\comsol}~\cite{Janssens:2890572}, employing FEM solvers (\methodB), and a spectral method (\methodC) that we have developed for this case.

\subsubsection[Time-dependent weighting potential computation with a FEM solver\\ (\methodB)]%
{Time-dependent weighting potential computation with a FEM solver (\methodB)}\label{sec:wfmapsFEM}
The \texttt{\comsol} software package implements a solver based on FEMs
which allows to solve the \qsme using the Backward Differentiation Formula (BDF)
method~\cite{Curtiss:1952, Gear:1967}. The toolkit allows to solve field equations by modeling geometries made from nearly arbitrary shapes.

Computations with FEMs rely on the process of dividing
the physical surfaces and volumes into a set of discrete two- or three-dimensional elements,  known as meshing.
These elements are designed to locally approximate the potential using second-order polynomials. 
A significant advantage of the meshing approach is that finer meshes can be designed near the 
electrodes, where the variations of the potential are expected to be more pronounced, 
while coarser meshes can be adopted in regions of slowly varying potential, thus optimizing the computational resources.

The computation of the time-dependent weighting potential is repeated for each electrode separately,
using the following procedure:
\begin{itemize}
    \item disregard any externally impressed or static charge densities;
    \item impose ideal grounding to all electrodes except the one under study, to which 
        a voltage ramp $V_w \tilde \Theta(t)$ is applied. We impose ideal grounding to all electrodes except the one under study, to which a voltage ramp 
$V_w \tilde{\Theta}(t)$ is applied. Here, $\tilde{\Theta}(t)$ is a smoothed version of the Heaviside step function, with a sub-femtosecond rise time $\tau$ chosen much shorter than the response time of the resistive material;
    \item evaluate the solution at logarithmic spaced time points to accurately capture the 
        early-time dynamics of the solution; as for spatial meshing, the time grid can be coarsened
        at large times to save computing resources, without significantly affecting the result. 
\end{itemize}

\begin{figure}
    \centering
    \includegraphics[width=0.48\textwidth]{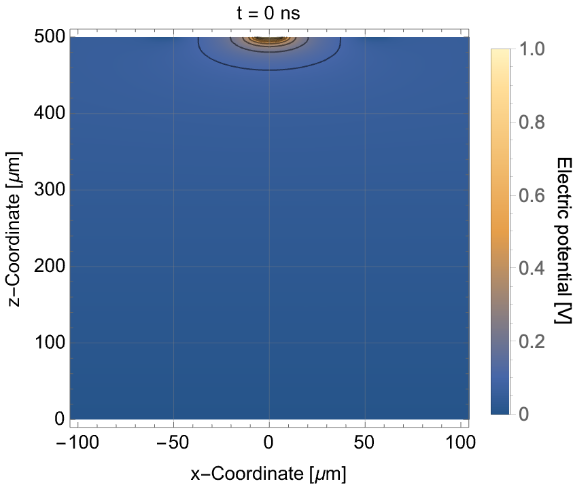}
    \includegraphics[width=0.48\textwidth]{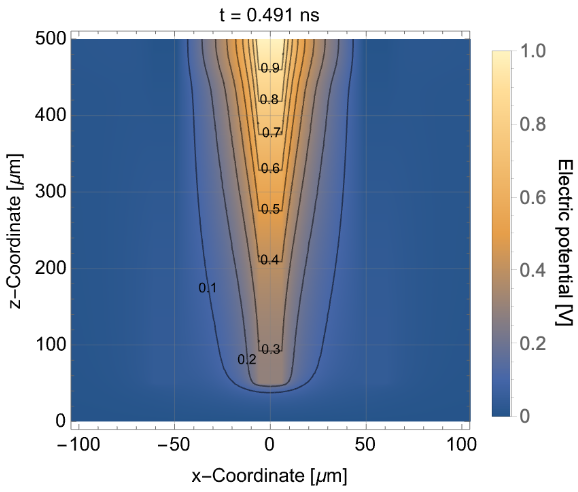}
    \includegraphics[width=0.48\textwidth]{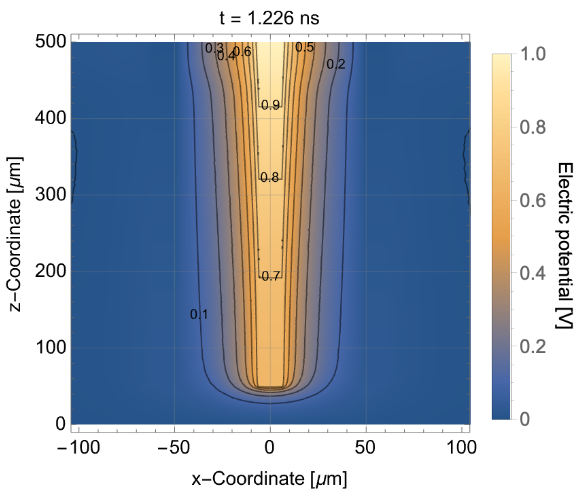}
    \includegraphics[width=0.48\textwidth]{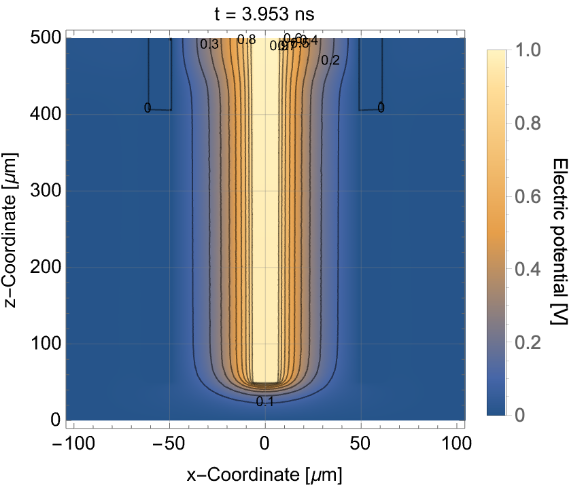}
    \caption{\label{fig:comsol-solution}
        Solution of the \qsme using the FEM as implemented in 
        \texttt{\comsol}. The four images represent the weighting potential of 
        a central readout electrode, for the sensor described at the beginning of Section 2, at four subsequent times following the voltage ramping up: 
        0, 0.5, 1.2 and 4.0~ns.  
        Figures are reproduced from Ref.~\cite{Janssens:2890572}.
    }
\end{figure}

Figure~\ref{fig:comsol-solution} depicts the solution of the \qsme for our considered geometry, as captured at four subsequent times 
following the initial voltage ramp. For the calculation the resistivity of the graphitized material is obtained assuming a resistance
of the  electrode of 30~$\mathrm{k\Omega}$. The relative dielectric constant of diamond is approximated as $\mathrm{\epsilon_r}$ = 5.7.

Once the approximated solution of the \qsme is known on the four-dimensional mesh, 
it is exported to a text-file which feeds \texttt{Garfield++} through a dedicated software component
named \texttt{Garfield::ComponentCOMSOL}, interpolating the solution on the mesh at arbitrary 
points and times as needed to compute the signal induced on the electrodes by the motion
of the carriers along their trajectories.

\subsubsection{Time-dependent weighting potential computation with spectral method (\methodC)}\label{sec:wfmapsFFT}
To provide a cross-check of the solution obtained with the FEM implemented by \texttt{COMSOL Multi-}

\noindent
\texttt{Physics\textsuperscript{\tiny\textregistered}\xspace}
and aiming at a fully open-source, possibly faster solver, we have implemented a custom algorithm 
based on the spectral method specifically intended for \qsme. 

The weighting potential $\Psi$ is described as the sum of two contributions: $\Psi_{\rho}$, representing 
the potential generated by the free charge density, and $\Psi_{ext}$, with $\nabla^2 \Psi_{ext}(\vec x, t) = 0$ 
almost everywhere, as will be described, used to enforce the boundary conditions. Hence,
\begin{equation}
    \label{eq:refactored-poisson}
    \nabla^2 \Psi(\vec x, t) = \nabla^2 (\Psi_{\rho}(\vec x, t) + \Psi_{ext}(\vec x, t)) = \nabla^2\Psi_{\rho}(\vec x, t) = - \frac{\rho}{\epsilon}.
\end{equation}
$\Psi_{\rho}$ and $\rho$ can be expanded as:
\begin{equation}
    \label{eq:expansion}
    \Psi_{\rho}(\vec x, t) = \sum_{j_x j_y j_z} c_{j_x j_y j_z} \exp\big(2\pi \ii \vec k \cdot \vec x\big) 
    \qquad \rho(\vec x, t) = \sum_{j_x j_y j_z} q_{j_x j_y j_z} \exp\big(2\pi \ii \vec k \cdot \vec x\big) 
\end{equation}
where $j_{x,y,z}$ are indices running from $-n_{x,y,z}/2$ to $n_{x,y,z}/2 - 1$, with $n_{x,y,z}$ being the number of uniform voxels in the mesh for the three spatial directions.

As customary with spectral methods, $\Psi_{\rho}$ is computed in voxels of the volume 
$\Delta x \times \Delta y \times \Delta z$ by noticing that
\begin{equation}
    \label{eq:spectralmethod}
    \nabla^2 \Psi_{\rho}(\vec x, t) = -\sum_{j_x j_y j_z} c_{j_x j_y j_z} (2\pi)^2 |\vec k|^2 \exp\big(2\pi \ii \vec k \cdot \vec x\big) = 
    -\frac{1}{\epsilon} \sum_{j_x j_y j_z} q_{j_x j_y j_z} \exp\big(2\pi \ii \vec k \cdot \vec x\big) = -\frac{\rho(\vec x, t)}{\epsilon},
\end{equation}
with $\vec k = \left({j_x}/{\Delta x}, {j_y}/{\Delta y}, {j_z}/{\Delta z}\right)$, which implies 
\begin{equation}
    \label{eq:solution}
     c_{j_x j_y j_z} = \frac{q_{j_x j_y j_z}}{(2 \pi)^2 |\vec k|^2\epsilon}. 
\end{equation}
The coefficients $q_{j_x j_y j_z}$ can be efficiently calculated by performing a Fast Fourier Transform (FFT) of the charge distribution $\rho(\vec{x},t)$. Once they are known, using the $c_{j_x j_y j_z}$ from Equation~\ref{eq:solution}, $\Psi_{\rho}$ is obtained from Equation~\ref{eq:expansion} via an inverse FFT.

The boundary conditions on the $x$ and $y$ axes (normal to the electrodes) are cyclic by construction. 
To break the periodicity along the $z$ axis, the sensor volume is padded with a fictional volume 
on which $\sigma(\vec{x})$ is defined in such a way as to ensure a smooth connection between the two oppositely polarized 
sides of the sensor. 

Additional boundary conditions are needed to constraint the voltage of the chosen readout electrode to $V_w$.
This is achieved by defining a finite number of points, named
voltage pins, corresponding only to
the electrode ends and defining $\Psi_{ext}$ as a weighted sum of the potential induced by point-like charges 
satisfying the boundary conditions on $\Psi_{\rho}+\Psi_{ext}$. 
This mechanism mimics the effect of an ideal voltage supplier injecting charge in a limited number of pre-defined point of the sensor to guarantee its setting conditions.

In practice, we define $\delta \psi (\vec x)$ the fundamental solution of the Poisson equation on the chosen 
discretization lattice for a unit charge positioned at the origin of the coordinate system.
The potential in $\vec x_1$ induced by a point-like charge $q$ placed in $\vec x_2$ is then 
\begin{equation}
    \label{eq:elementarysolution}
    \Psi_{ext}(\vec{x_1},t) = \delta\psi(\vec x_1 - \vec x_2) q.
\end{equation}

Considering $N$ voltage pins placed in $\vec x_{p_1}, \vec x_{p_2}, ... \vec x_{p_N}$, we can compute 
the total potential in each position as the sum of $\Psi_{\rho}$ and of the $\Psi_{ext}$, as defined by the voltage pins themselves,
\begin{equation}
    \label{eq:voltagepins-extended}
    \left(
        \begin{array}{c}
        \Psi_{p_1} \\
        \Psi_{p_2} \\
        \vdots \\
        \Psi_{p_N} \\
        \end{array} 
    \right) = \left( 
        \begin{array}{cccc}
        \delta \psi(\vec x_{p_1} - \vec x_{p_1}) & \delta \psi(\vec x_{p_1} - \vec x_{p_2}) & \cdots & \delta \psi(\vec x_{p_1} - \vec x_{p_N}) \\
        \delta \psi(\vec x_{p_2} - \vec x_{p_1}) & \delta \psi(\vec x_{p_2} - \vec x_{p_2}) & \cdots & \delta \psi(\vec x_{p_2} - \vec x_{p_N}) \\
        \vdots & \vdots & \ddots & \vdots \\
        \delta \psi(\vec x_{p_N} - \vec x_{p_1}) & \delta \psi(\vec x_{p_N} - \vec x_{p_2}) & \cdots & \delta \psi(\vec x_{p_N} - \vec x_{p_N}) \\
        \end{array}
    \right)
    \left(
        \begin{array}{c}
        q_{p_1} \\
        q_{p_2} \\
        \vdots \\
        q_{p_N} \\
        \end{array}
    \right)
    +
    \left(
        \begin{array}{c}
        \Psi_{\rho} (\vec x_{p_1}) \\
        \Psi_{\rho} (\vec x_{p_2}) \\
        \vdots\\
        \Psi_{\rho} (\vec x_{p_N}) \\
        \end{array}
    \right)
\end{equation}
or, in short,
\begin{equation}
    \label{eq:voltagepins-contracted}
    \mathbf {\Psi_{\vec{\rm p}}} =  \mathbb{D} \mathbf{q} + \mathbf{\Psi_{\rho}} 
\end{equation}
which implies that the charge $\mathbf q$ requested to constrain the voltage at the chosen 
$\mathbf{\Psi_{set}}$ can be computed as
\begin{equation}
    \label{eq:voltagepingcharge}
    \mathbf q(t) = - \mathbb D^{-1} \mathbf{\Psi_{\rho}}(t) + \mathbb D^{-1} \mathbf {\Psi_{set}},
\end{equation}
where the dependence on the time coordinate has been made explicit. 
Note that $\mathbb D$ and $\mathbf{\Psi_{set}}$ can be considered constants as
neither the position of the voltage pins nor the corresponding voltage set points are supposed 
to evolve during the simulation.
Hence, 
\begin{equation}
    \label{eq:vext}
    \Psi_{ext} = \sum_{i=1}^{N} q_{p_i}(t) \delta \psi(\vec x - \vec x_{p_i}).  
\end{equation}

The time evolution of the potential follows the evolution of the charge density $\rho(\vec x, t)$
as described by the continuity equation
\begin{equation}
    \label{eq:continuity}
    \frac{\partial \rho (\vec x, t)}{\partial t} - \vec \nabla \cdot (\sigma (\vec x) \vec \nabla \Psi (\vec x, t)) = 0.
\end{equation}
Following the prescriptions of the \emph{forward Euler} method, and adopting the spectral method to approximate 
the derivatives of $\Psi_{\rho}$, we write the evolution step as 
\begin{equation}
    \label{eq:eulerstep}
    \rho(\vec x, t + \mathrm d t) = \rho(\vec x, t) 
        - \mathcal F^{-1} \left( 
            2\pi\vec k \cdot \mathcal F \left(
                \sigma(x)\mathcal F^{-1} \left(
                    2\pi \vec k \mathcal F(\Psi(\vec x, t))
                \right)
            \right)
        \right)\ \mathrm d t,
\end{equation}
where $\mathcal F$ and $\mathcal F^{-1}$ indicate a Fourier Transform and its inverse, respectively, 
and are implemented using the FFT module of the \texttt{JAX} library~\cite{jax2018github}.

\begin{figure}
    \centering
    \includegraphics[height=0.9\textheight]{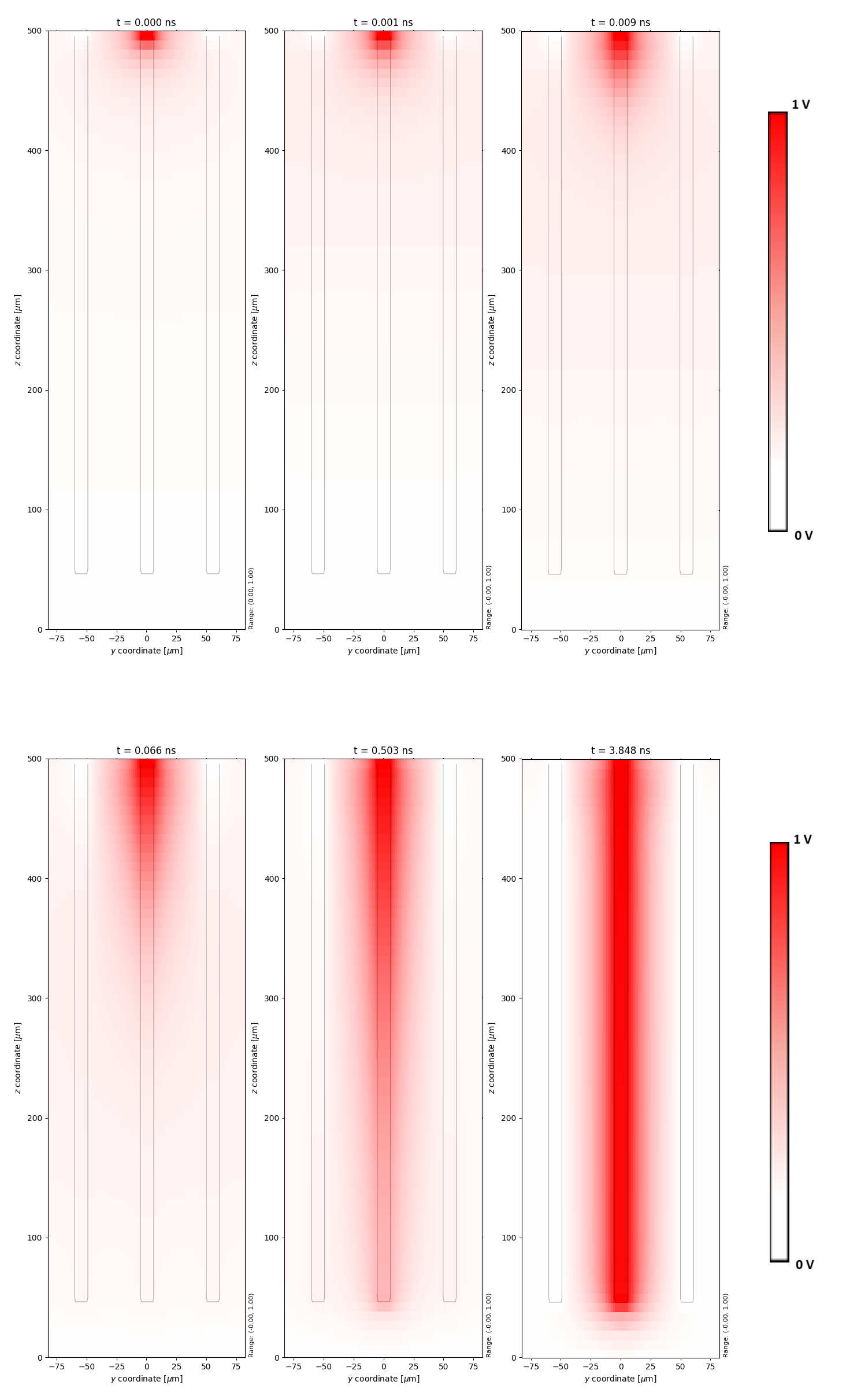}
    \caption{
        \label{fig:fft-solution-columnar-sensor}
        Solution of the \qsme using the spectral method described in text.
        The six images represent the weighting potential of 
        a central readout electrode, for the sensor described at the beginning of Section 2, at 0, 0.001, 0.009, 0.07, 0.5 and 3.8~ns.  
     }
\end{figure}

The weighting potential obtained with this method for the geometry of the sensor under study using a mesh of $120 \times 120 \times 64$ voxels is shown in Figure~\ref{fig:fft-solution-columnar-sensor} at six different times.

The solutions derived from the \qsme using the spectral method are 
transferred into ASCII-formatted files. These files are subsequently 
processed by \texttt{Garfield++} through its standard component \texttt{Garfield::ComponentVoxel},
which implements the interpolation of time-dependent weighting potentials in
a manner similar to \texttt{Garfield::\allowbreak ComponentCOMSOL}.

\subsection{Comparison of signal simulation methods}
\label{sec:comparison}

In the new simulation approach, the effect of resistance is modeled through time-dependent weighting potential maps, computed with either \texttt{\comsol} (\methodB) or the spectral solver (\methodC). These maps are processed by \texttt{Garfield++} implementing the convolution between 
the time evolution of the induced current due to the carrier transport and that due to 
the signal propagation as described in Equation~\ref{eq:ramoshockleyriegler}.

\begin{figure}
    \begin{minipage}[t]{0.46\textwidth}
        \includegraphics[width=1.0\textwidth, trim=20mm 0 0 0]{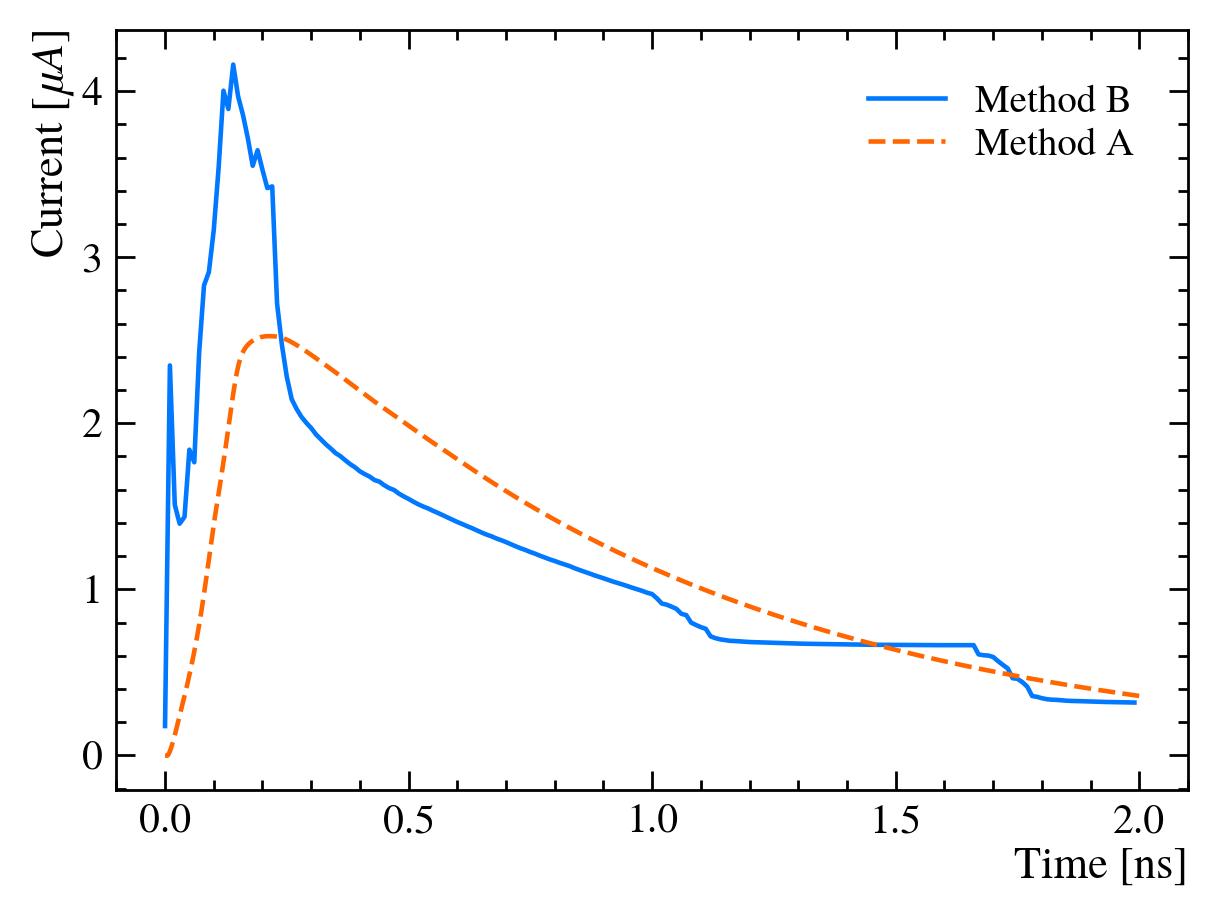}
        \caption{\label{fig:confrontospice}
           Signals induced on resistive electrodes embedded in a diamond 
        dielectric traversed by a 180 GeV/$c$ pion. Comparison between the signals obtained from a complete simulation with time-dependent 
            weighting potential (blue, \methodB) and by convolving with the discrete impedance network 
            transfer function with the simulation for the metallic electrodes obtained with
            a steady weighting potential (orange, \methodA).
        }
    \end{minipage}
    \hfill
    \begin{minipage}[t]{0.48\textwidth}
        \includegraphics[width=1.0\textwidth, trim=20mm 0 0 0]{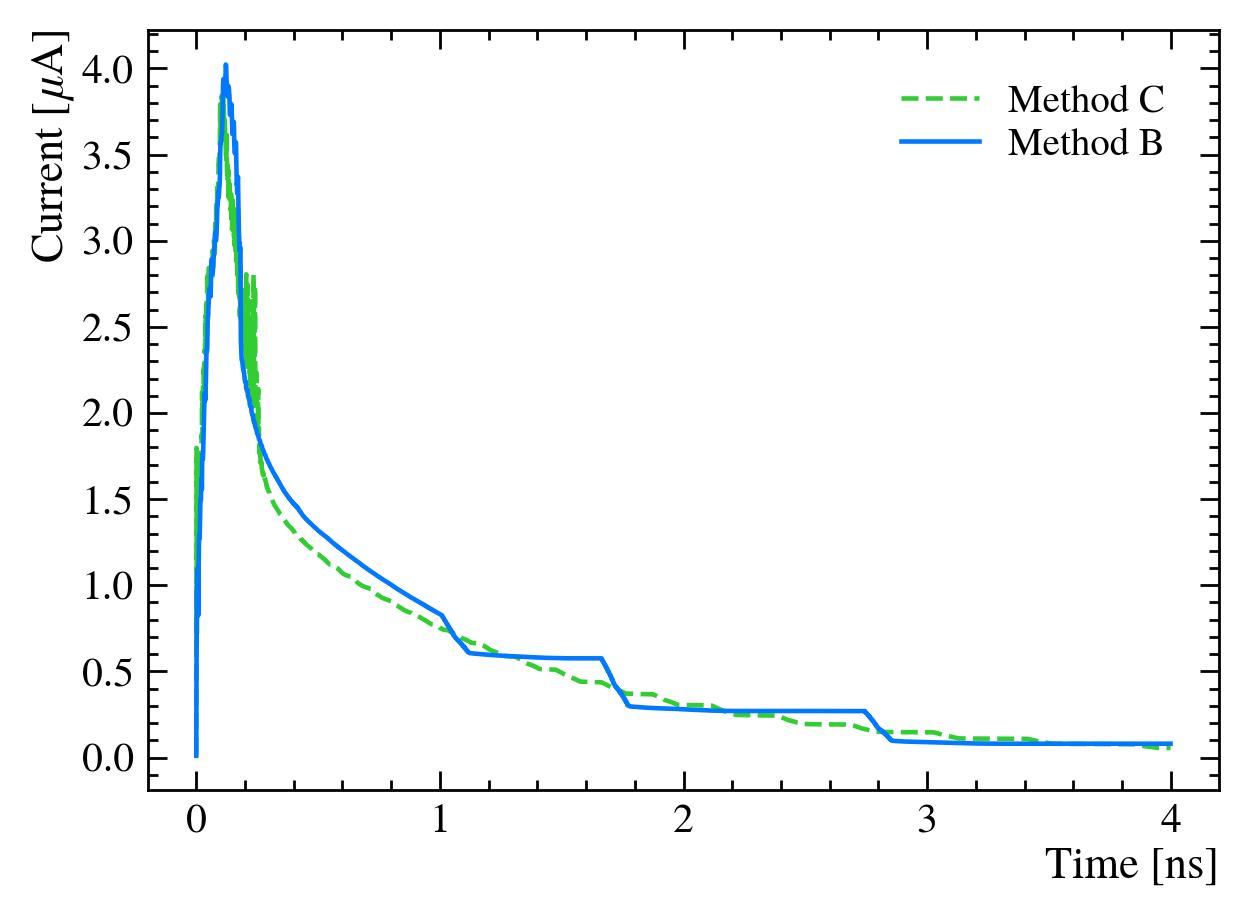}
        \caption{\label{fig:fftvscomsol}
            Signals induced on resistive electrodes embedded in a diamond 
        dielectric traversed by a 180 GeV/$c$ pion. Comparison between the signals obtained from a complete simulation with 
            time-dependent weighting potentials as computed with a FEM
            implemented in \texttt{\comsol} (blue, \methodB) and with our custom solver 
            based on spectral methods (green, \methodC).
        }
    \end{minipage}
\end{figure}

The current induced on the considered readout electrode using the maps obtained with the FEM 
implementation in \texttt{\comsol} (\methodB) is compared to the simplified method 
representing the graphitic electrodes as discrete impedance networks (\methodA) in Figure~\ref{fig:confrontospice}. 
The comparison is performed using the same simulated 180 GeV/$c$ pion, with identical trajectory and 
ionization energy deposits, and the same electrostatic field map determining the 
trajectories of the charge carriers. 
The two models differ significantly in the early stages of the signal development, where the simple impedance network fails to capture the current induction close to the readout electronics input during the evolution of the time-dependent weighting field.
Indeed, once the charge carriers have been all collected and the current is dominated by the delayed component due to the signal propagation (process 3) through the electrodes, the predictions of the two models are consistent. 

The disagreement between the predictions in the first part of the signal, corresponding to the time interval where process 2 and process 3 coexist, raises concerns about the reliability of \methodA at lower electrode resistances. 
Process 3 becomes negligible in the limit of metallic electrodes, where it is effectively instantaneous, and the agreement between models is ensured by the identical treatment of process 2. However, there is no evidence that the simplified model correctly captures the coupled dynamics of processes 2 and 3 when the resistance is finite. This limitation becomes particularly relevant when considering optimization strategies for detector geometry or readout electronics, as the early signal region, where the discrepancy is most pronounced, plays an important role in determining the final time resolution.

The signals obtained from another incident particle using the FEM (\methodB) and
the spectral methods (\methodC) are compared in Figure ~\ref{fig:fftvscomsol}.
The FEM signal,
derived from a variable size mesh, adapted to the geometry, is smoother than that obtained with
the spectral method on a uniform voxel matrix. The spikes produced by
the spectral method are indeed artifacts due to the rapid variation
of the potential in the proximity of the electrodes which is not well
represented by the fixed size voxels. This problem could be mitigated
with a finer voxel pitch at the price of a considerably longer
computing time. Nevertheless, the two signals agree very well and the
computing artifacts are washed out once the raw signal is convoluted
with any realistic readout electronics response function as discussed
in the next section.

\subsection{Effects of readout electronics and electronic noise}\label{sec:kuboard}

In order to compare the simulation results with the data collected during the
beam test described in Ref.~\cite{Anderlini:2022exf}, it is essential to properly take into account the effects due to the signal processing in the various steps of the readout. This is obtained by convolving the signal obtained from the sensor simulation described in the previous sections with the transfer function of the readout electronics, and properly taking into account the corresponding electronic noise. For a full comparison, features of the data acquisition chain such as digitalization sampling frequency  will also be considered.

The readout front-end board employed in the beam test was developed at the 
University of Kansas (referred to as the "KU board" in the following)~\cite{Minafra:2017nqc}.
The KU board is based  on a low cost two-stage common-emitter amplifier built with fast Si-Ge transistors. The board, originally developed for Ultra-Fast Silicon detectors, has been adapted to diamond sensors by increasing the gain of the amplifier to deal with the lower amount of charge released in diamond. This goes at the expenses of the timing performance which was not optimal for the 3D sensors used in the test.

The output signals of the KU board were acquired through a 6~GHz digital oscilloscope at a sampling rate of 20~GS/s, saving the waveforms on disk. Therefore, to properly simulate the full readout chain, the simulated detector signals were convolved with the KU board transfer function and interpolated at 20~GS/s. We considered the effect of the analog 6~GHz oscilloscope bandwidth negligible compared with the KU board bandwidth.

\begin{figure}
    \centering
    \includegraphics[width=0.6\textwidth]{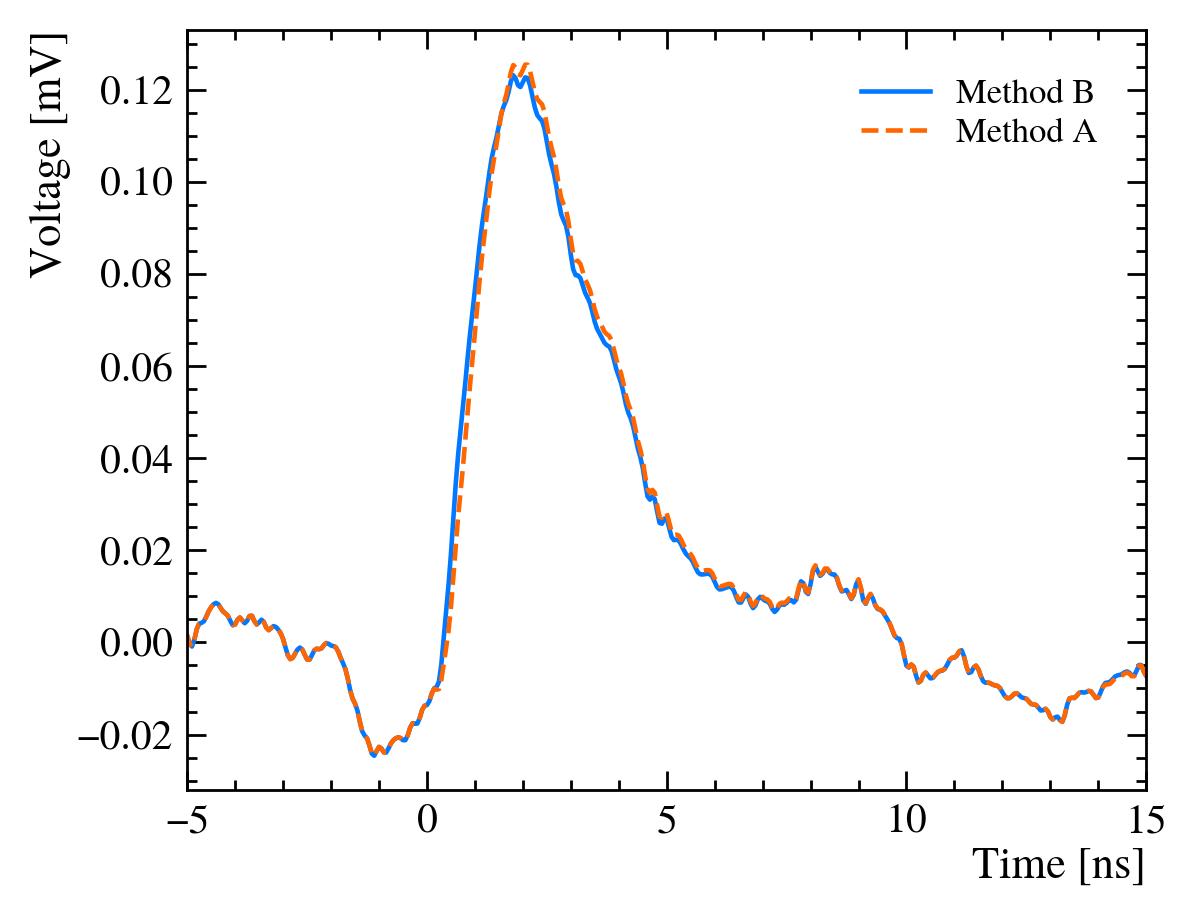}
    \caption{
        \label{fig:comsol-vs-impedance-net-single-signal}
        Comparison between the waveforms obtained with the \texttt{Garfield++} simulation
        using weighting potentials obtained from \texttt{\comsol} (blue, \methodB) and 
        with the steady simulation convolved with the transfer function 
        of the discrete impedance network modeling the effects of 
        signal propagation (orange, \methodA).
        The noise contribution, superposed \emph{a posteriori}, is identical for the two signals.
    }
\end{figure}

Since the KU board has a non-trivial noise spectrum, in the simulation we empirically summed noise waveforms acquired with the oscilloscope in auto-trigger mode, with the sensor connected and polarized,  to the simulated waveforms.
An example of waveform obtained through the whole simulation workflow is shown 
in Figure~\ref{fig:comsol-vs-impedance-net-single-signal}, where the results 
obtained with impedance network and time-dependent weighting potentials are superposed. 

At the end of the simulation flow discussed so far, both the experimental and simulated waveforms can be compared by extracting parameters like their amplitude, rise-time and delay with respect to a reference timing signal to measure the time resolution.
The latter can be measured using for example a digital constant 
fraction discriminator algorithm, as described in detail in Ref.~\cite{chiara}. 

\begin{figure}
    \centering
    \includegraphics[width=0.48\textwidth]{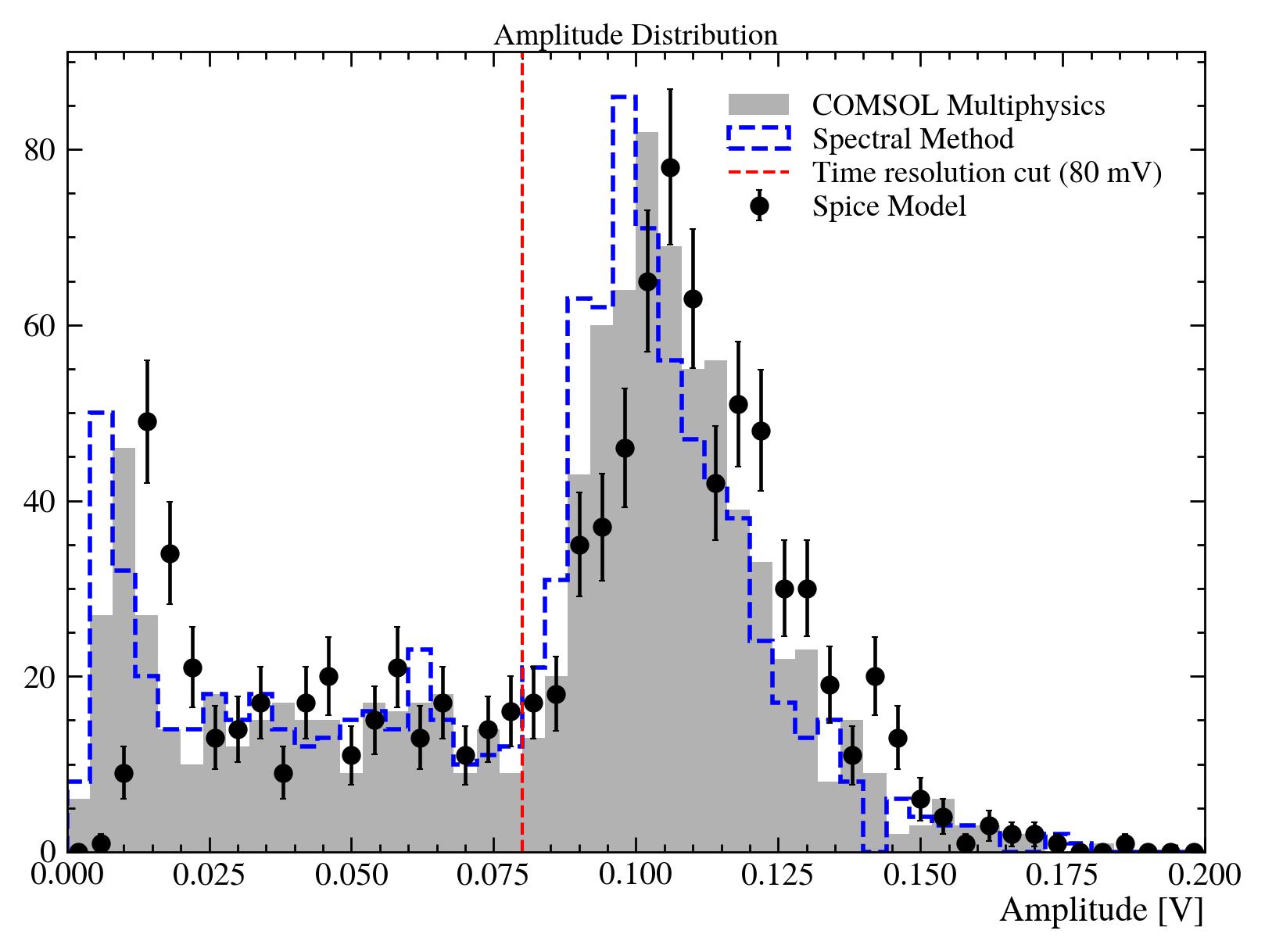}
    \includegraphics[width=0.48\textwidth]{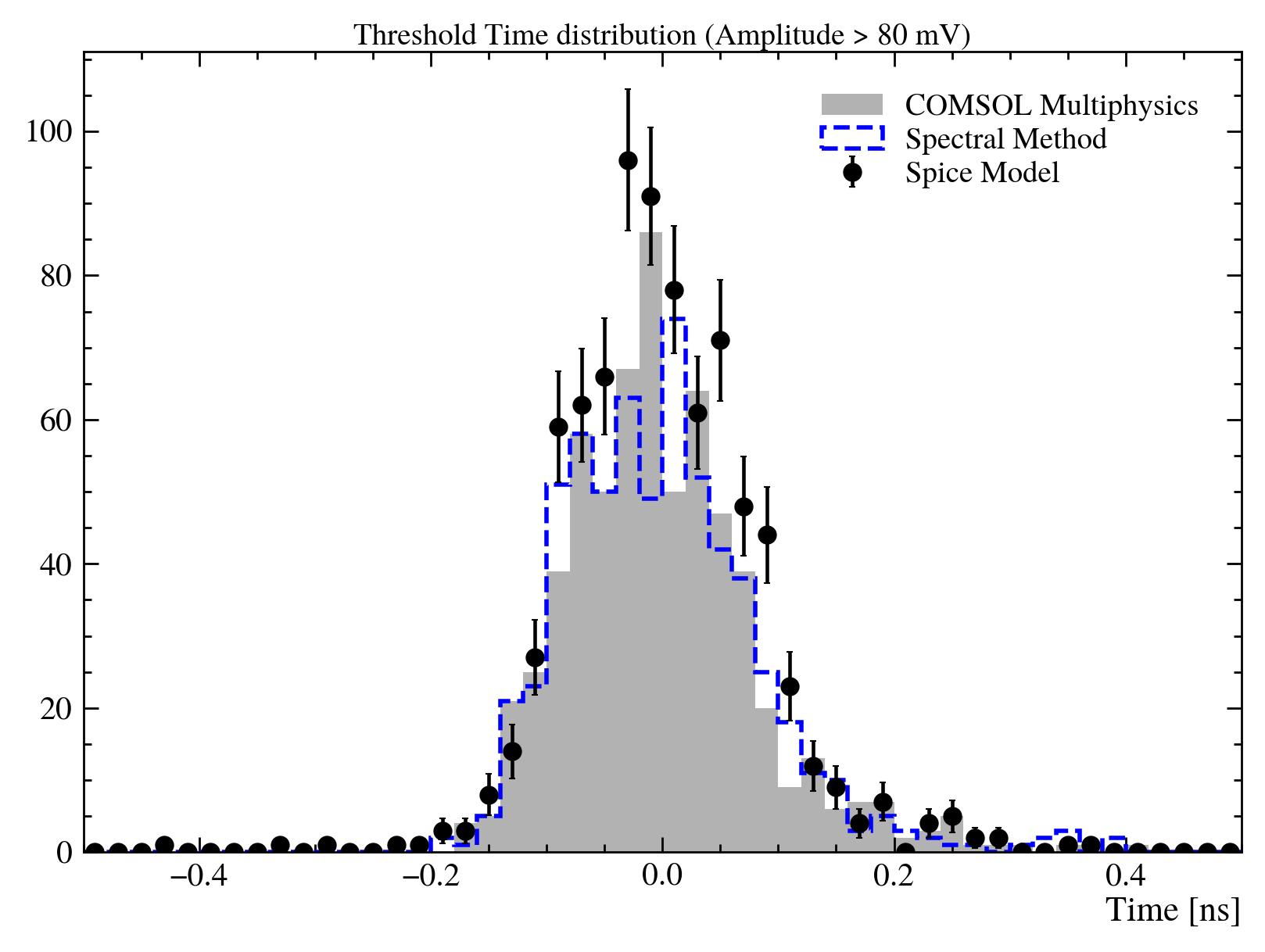}
    \caption{
        \label{fig:comsol-vs-fft-with-electronics}
        Comparison of the distributions of amplitudes (left) 
        and of time markers (right) of signals from the full simulation chain described in the text, obtained with the discrete impedance network (black markers with error bars, \methodA) and with time-dependent weighting 
        potentials calculated with \texttt{\comsol} (gray filled, \methodB) and  
        with the spectral method (blue dashed line, \methodC).
    }
\end{figure}

The distributions of the simulated signal amplitudes and time markers, utilizing the time-dependent 
weighting potential as computed with the FEM solver (\methodB) and the spectral method (\methodC), and obtained with the discrete impedance network (\methodA) are presented in Figure~\ref{fig:comsol-vs-fft-with-electronics}. 
Within the limits of the spectral method that must be refined as discussed in the previous sections, we consider the results obtained with \methodB and \methodC as consistent in the description of the signal amplitude spectrum. The simple discrete impedance description, besides being ad hoc, tends to slightly overestimate the signal amplitudes but is qualitatively consistent with the two previous methods. The time distribution of the signal is largely insensitive to the underlying induced current distributions, as it is dominated by the shaping effects of the readout electronics. Consistent results are obtained regardless of whether the weighting potential is computed using spectral or FEM solvers.

We conclude that both \methodB and \methodC provide predictions consistent with the heuristic \methodA in the regime where it is expected to be reliable, and consistent with each other, especially once the amplification and shaping effects (process 4) are taken into account.
Hence, we consider \methodB and \methodC as sufficiently robust to allow a direct comparison with experimental data, as discussed in the next section.

\section{Validation of simulation on beam-test data}
\label{sec:data}

\begin{figure}
    \centering
    \includegraphics[width=0.8\linewidth]{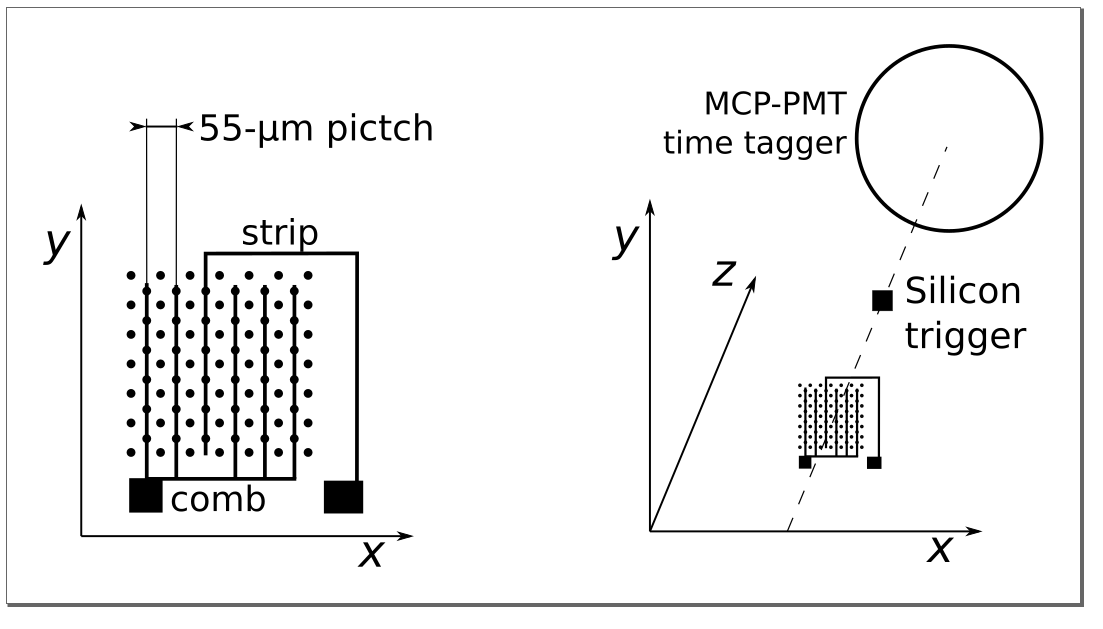}
    \caption{
        \label{fig:test-beam}
        On the left, a view of the sensor under study during the beam test. 
        On the right, the schematic of the experimental setup used at the beam test.
    }
\end{figure}

To validate the simulation chain described above, 
we perform a comparison with data obtained during a beam test at CERN SPS in 
2021~\cite{Anderlini:2022exf}. 
The tested device was a prototype $6\times 6$ 3D pixel matrix with $55\times 55\mu$m$^2$ pitch and a geometry identical to that described in Section~\ref{sec:sim}.
The diameter of the electrodes, measured with an optical microscope from the shadow projected by the nano-structured 
graphitized material, is $\sim$12~$\mu$m.
As the electrodes terminate with only one end at the diamond crystal surface (see Figure~\ref{fig:schematic-view}), direct measurement of their resistance was not feasible. An indirect evaluation was therefore performed using specially fabricated graphitized through-columns, in which both ends were accessible. These reference electrodes were produced on the same diamond crystal, within the same fabrication run, and under identical process parameters as those adopted for the sensor. On this basis, the resistance of the sensor electrodes was estimated to be approximately 30~k$\Omega$.
The electrode fabrication parameters were carefully optimized to lower as much as possible the resistance but the actual optical and lattice conditions may worsen
the quality of the graphitization, making the measured 30~k$\Omega$ more a lower limit than an expected value.

For the beam test, to enhance the active area, several pixels were connected in parallel into a strip of six cells and a comb-like arrangement of thirty cells by means of surface graphitic traces, as illustrated in Figure~\ref{fig:test-beam}. To select the beam particles traversing the active area (or equivalently, to \emph{veto} particles at the boundaries of the sensor or beyond), we used a single-pixel silicon sensor with an active area of $55\times 55$~$\mu$m$^2$~\cite{Anderlini:2020ffm} combined with a Micro-Channel Plate Photomultiplier Tube (MCP-PMT) which provided a trigger signal with a timing precision better than 10~ps~\cite{Garau:2023fyo}.
The position of the silicon pixel relative to the diamond was adjusted using a piezoelectric linear stage in order to allow a scan of the diamond matrix active area. The signals were read out and acquired through the electronics chain described in Section~\ref{sec:kuboard}.

To compare the simulation with experimental data, we use the signals from the six-pixel strip as a proxy of a single pixel. It has to be noted that the strip is not exactly equivalent to a single pixel and some charge is lost through the pixels connected in parallel, due to their finite impedance, before reaching the readout electronics input. The comparison focuses on two main observables: the distribution of signal 
amplitudes, defined as the maximum value of each waveform after baseline subtraction, and the 
distribution of time differences (\emph{delays}) between the diamond signals and the reference MCP-PMT signal, from which the time resolution is extracted. 

\begin{figure}[t]
    \centering
    \begin{minipage}[t]{0.48\textwidth}
        \centering
        \includegraphics[width=\linewidth]{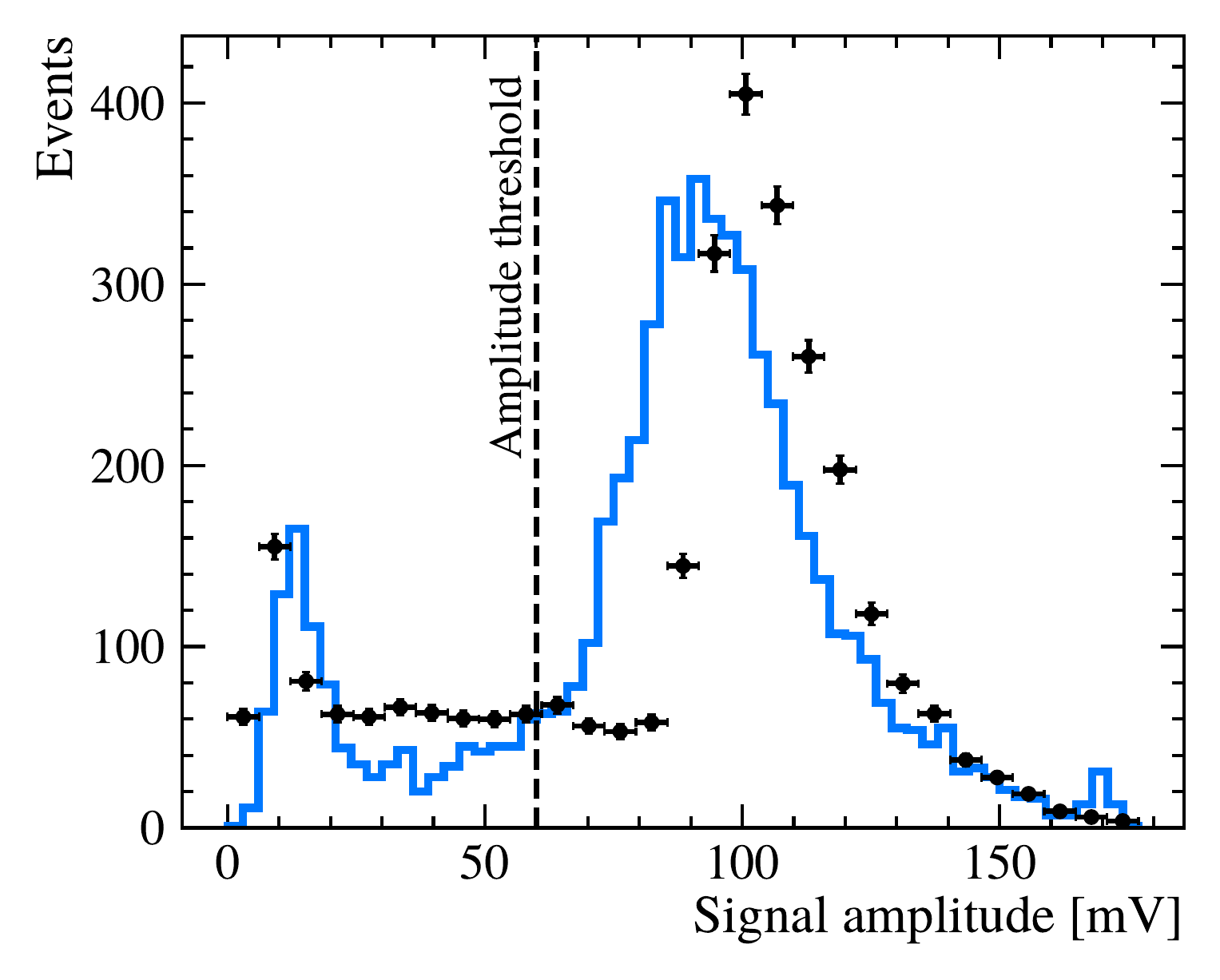}
        \caption{
            \label{fig:amplitude-comparison}
            Distributions of signal amplitudes as obtained analyzing 
            data acquired during the beam test at the
            CERN SPS in 2021 (blue solid line); 
            and from 10000 simulated tracks using field maps generated with \texttt{\comsol} for 30~k$\Omega$
            electrodes, vertically rescaled to match data normalization (markers with error bars).
        }
    \end{minipage}
    \hfill
    \begin{minipage}[t]{0.48\textwidth}
        \centering
        \includegraphics[width=\linewidth]{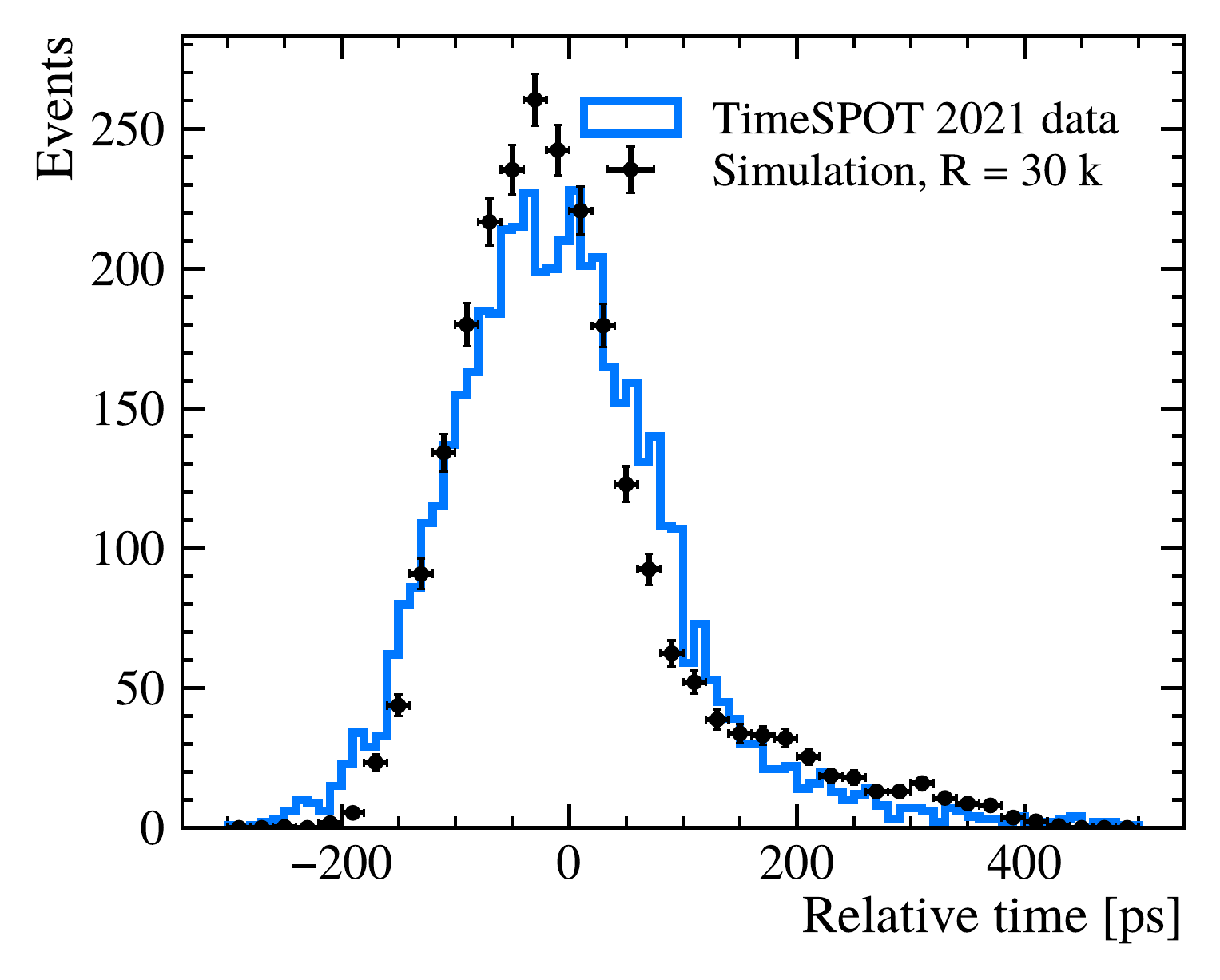}
        \caption{
            \label{fig:timing-comparison}
            Distributions of timing measurements as obtained analyzing events with 
            an amplitude larger than 60~mV in the 
            data acquired during the beam test at the
            CERN SPS in 2021 (blue solid line); 
            and from 10000 simulated tracks with weighting potentials
            generated with \texttt{\comsol} for 30~k$\Omega$ electrodes, 
            vertically rescaled to match data normalization (markers with error bars).
            }
    \end{minipage}
\end{figure}

Figure~\ref{fig:amplitude-comparison} shows the amplitude distributions 
obtained from the simulation and the test beam data. The two distributions exhibit 
a similar structure and share common features. 

In the simulation, a peak near 0 V is observed, which can be attributed to 
three main categories of events: events where both end points of the particle 
trajectory lie within the readout electrode, events with the final end-point 
outside the simulation volume, and events occurring in proximity to 
either of these categories. Beyond this peak, the simulated distribution 
displays the expected Landau shape. Between the two peaks of the distribution, 
there is a number of events associated with the presence of inclined tracks, where part of the charge is created outside the cell volume. A 
Landau fit of the main peak yields a most probable amplitude for the simulated amplitudes of 
$\mathrm{(101.7\pm0.6)~mV}$

The experimental amplitude distribution presents three similar components: an 
initial peak near 0 V, primarily due to low-amplitude signals from particles 
traversing the graphite columns, charge sharing with adjacent pixels, and 
electronic noise; a valley, again associated with inclined tracks; a Landau 
peak, whose most probable value is found at $\mathrm{(89.7\pm0.4)~mV}$. 

The 10\% discrepancy with respect to the prediction from simulation is consistent 
with an increased detector capacity due to the multiple electrodes connected to form 
the strip, but could also be due to an underestimation of the electrode 
resistance in the simulation setup, or even to a sub-fluctuation of the 
amplifier gain due to fabrication tolerances on the transistors.

The time resolution of both the simulated and experimental device is obtained from the distribution of the signal delays with respect to a reference time signal. For the simulated events, the reference time is conventionally set to zero, while for the real data this is provided by the MCP-PMT detector. The observed time distribution is not Gaussian and shows a significant tail at large relative times, due to several effects which, however, can be of different origin. 

In data, the long relative time tail is populated by low amplitude signals coming for example from the noise distribution tails or due to charge sharing between adjacent pixels. 
In simulation, instead, a significant number of events in the upper tail of the distribution are due to artifacts arising from the inaccurate modeling of charge carrier motion inside the electrodes.

The relative weight of the tail can thus be reduced by rejecting low-amplitude events. In Figure~\ref{fig:timing-comparison}, simulation and data are compared for signals with an amplitude larger than 60 mV. The two distributions are still slightly asymmetric although the central Gaussian core has a significantly larger weight. To properly take into account the residual distribution asymmetry, a fit is performed with a Crystal Ball function which features a Gaussian core and an exponential tail. The Gaussian core can be conventionally used to estimate the time resolution which is measured to be $(71 \pm 3)$~ps in the simulated events and  $(82 \pm 2)$~ps in the experimental data. 
The discrepancy is largely due to the differences in the signal amplitude spectrum discussed above. 

\medskip
Considering the differences between the simulated system and the experimental device, as well as the large uncertainty on the resistance of the graphitized electrodes, the agreement between data and simulation can be considered satisfactory. 
For example, repeating the simulation with a resistance of 60~k$\Omega$ for the electrodes, the predicted amplitude peak becomes $(76.9 \pm 0.8)$~mV. The corresponding time resolution is $(90 \pm 2)$~ps.
Hence, we considered the simulation procedure validated and ready to
guide the optimization of detector performance 
through targeted modifications in sensor fabrication techniques.

\section{Systematic detector studies based on simulations}
The simulation process illustrated in Section~\ref{sec:sim} and compared with real data in Section~\ref{sec:data} can be now used to systematically explore some of the critical parameters of the 3D diamond detectors to seek for possible optimizations. In particular, we want to systematically study the parameters that affect their time resolution in order to get indications for possible design improvements.

\subsection{Effect of electrode resistance and optimization}\label{sec:resistance}
From the comparison between results obtained with metallic electrodes 
and real resistive electrodes, it is evident that the resistance of the graphitized electrodes represents the dominant contribution to the time resolution and is thus the main construction parameter to be optimized.

The most direct effect of high electrode 
resistance is a proportional increase in the signal time constant, resulting in a 
longer rise time (${t_{\mathrm{rise}}}$), and to a lesser extent, fall time ($t_{\mathrm{fall}}$). 
Since the total collected charge, and hence the integral of the signal, remains constant, 
longer rise (and fall) time leads to a reduction in the signal amplitude (${v_{\mathrm{max}}}$). 
Assuming a constant noise level, the signal-to-noise ratio (SNR) decreases accordingly. 
The rise time and SNR determine the jitter contribution to the time resolution, given by:
\begin{equation}
    \label{eq:jitter}
    \sigma_{\rm jitter} = \frac{t_{\rm rise}}{\mathrm{SNR}} = t_{\rm rise}\frac{\sigma_v}{v_{\rm max}}
\end{equation}
where $\sigma_v$ is the standard deviation of the noise distribution.
Thus, an increased electrode resistance directly results in a larger
jitter contribution. From Equation~\ref{eq:jitter} we can estimate what would be the
minimum jitter contribution, with perfectly
conductive electrodes.
Convolving a very short current spike with the KU board
transfer function, resulting in a waveform with $t_{\mathrm{rise}} \approx 0.5\ \mathrm{ns}$ and $t_{\mathrm{fall}} \approx 4\ \mathrm{ns}$, and using the measured electronic noise (5 mV RMS) for $\sigma_v$,
we obtain a minimum jitter contribution of about 11~ps. 
Taking into account the uncertainty caused by variation in 
the drift times of charge carriers across a half cell, 
we estimate a minimum jitter contribution to the time 
resolution of approximately 15~ps.

\begin{figure}
    \centering
    \includegraphics[width=0.65\linewidth]{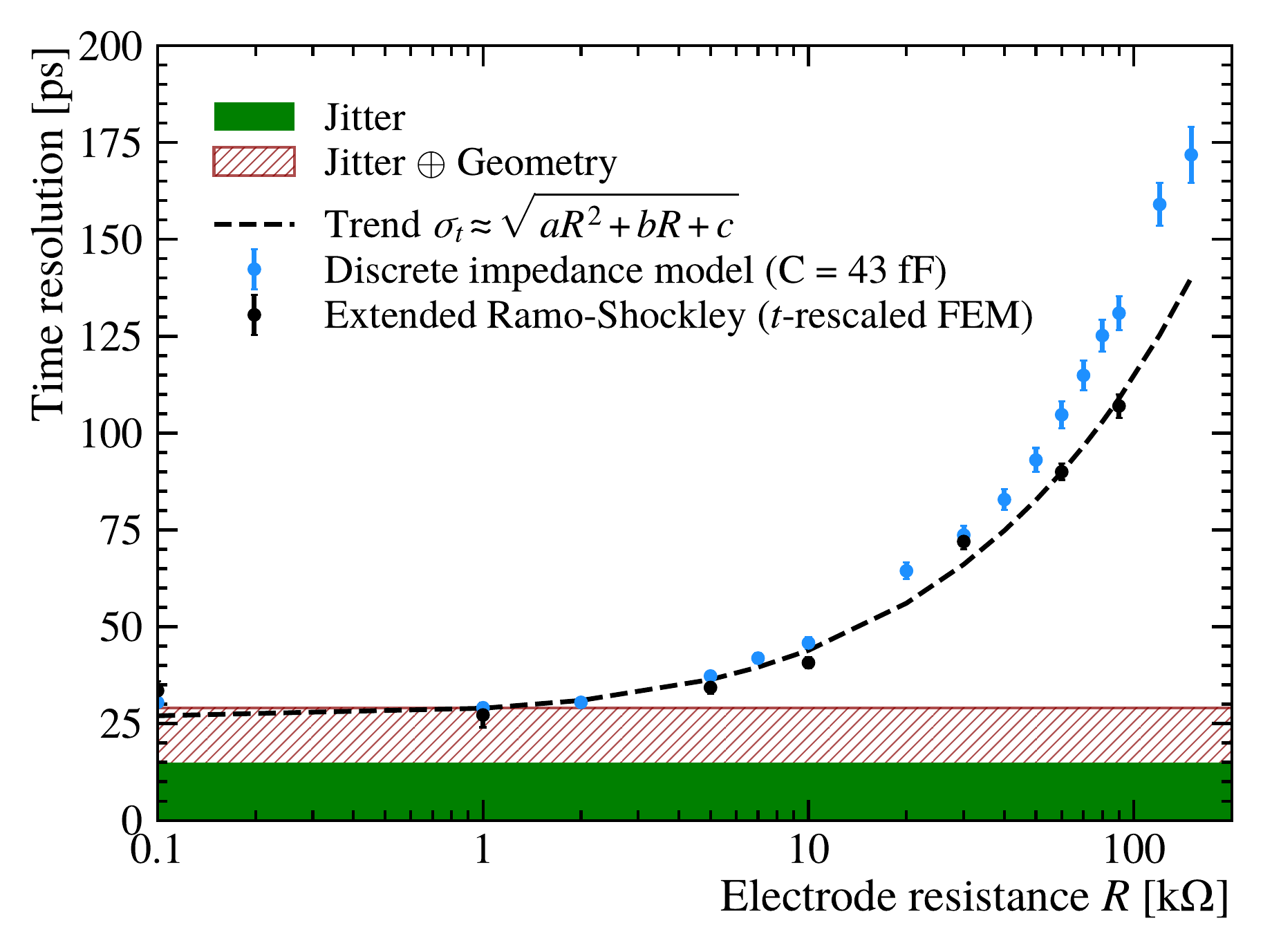}
    \caption{
        Time resolution as a function of electrode resistance ($R$). Blue markers 
        represent results from the discrete impedance network model (\methodA), whereas black markers show 
        \texttt{Garfield++} simulation results with weighting potentials computed with \texttt{\comsol} (\methodB) and
        rescaling the time axis by R/30 k$\Omega$. 
        A fitted curve (black dashed) guides the reader's eye.
        The expected contribution, independent on the electrode resistance, from electronic 
        noise and inhomogeneities of the electric field are also shown as green-solid and
        red-dashed bands.
        }
    \label{fig:cellmodel}
\end{figure}

With the simulation, we have studied in a systematic way the effect of the electrode resistance on the time resolution. The results of this study are reported in Figure~\ref{fig:cellmodel}. This has been achieved by changing the resistance in the discrete impedance model and by scaling the time dependence of the weighting potential (see Section~\ref{sec:ramo}).

While showing a significant discrepancy at large resistances (primarily due to the over-simplification of \methodA, especially in the early rise of the signal which is more relevant for time resolution) the two approaches converge to the same asymptotic time resolution below reasonably achievable resistance values of a few tens of k$\Omega$. 
The asymptotic minimum time resolution is given by the sum in quadrature of the jitter contribution described above and the contribution of the inhomogeneities of the electric field due to the sensor geometry that will be discussed in the next section.
The agreement at low resistance is not surprising, as process 3, the only one treated differently in \methodA and \methodB, becomes progressively less relevant in that regime.

Interestingly, the study indicates that for the sensor under study, with the described geometry and readout electronics, the effect of electrode resistance on time resolution becomes comparable or sub-dominant with respect to other contributions for resistance values below about 10 k$\Omega$. 

\subsection{Optimization of the geometry}
\label{sec:other-geometries}

The findings from the preceding sections demonstrate that in 3D diamond sensors, 
the time resolution is primarily influenced by signal propagation effects 
resulting from high electrode resistance. 
However, with the decreasing electrode resistance obtained by recent promising advancements in graphitization techniques,  non-uniformities in the electric field start to significantly contribute to the sensor timing performance. 
Consequently, exploring alternative geometries aimed at enhancing electric field 
uniformity becomes crucial for further improving time resolution.

Using the spectral method (\methodC),
modifying the sensor geometry and recomputing time-dependent weighting 
potential and electrostatic field maps is relatively fast and straightforward.
The generated maps can then be fed to \texttt{Garfield++} without further modifications
to the simulation, signal processing and data analysis flow.
\begin{figure}
    \centering
    \includegraphics[width=1\linewidth]{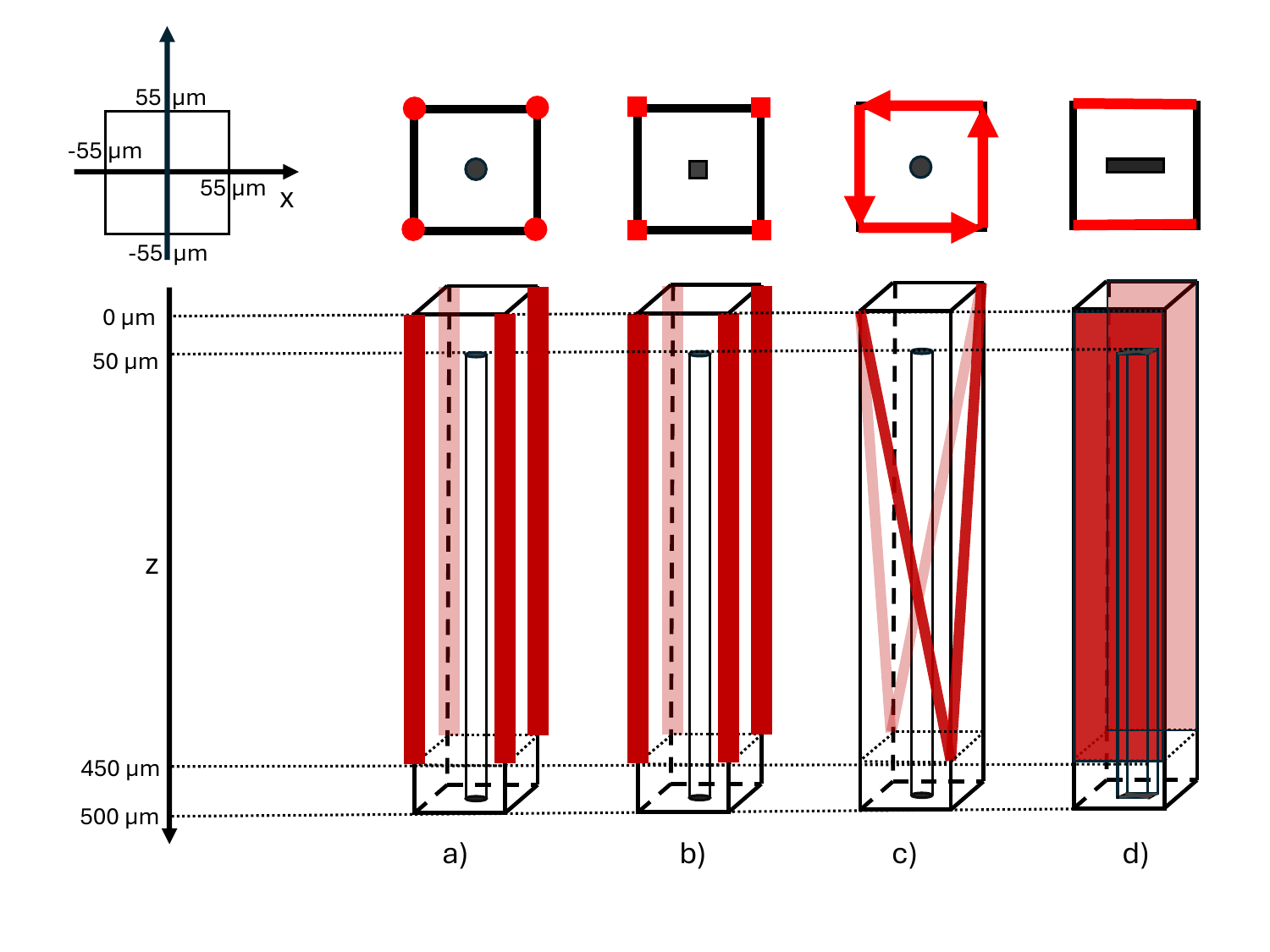}
    \caption{
        Cross section and prospective representation of all the tested geometries. Polarization electrodes are represented in red: a) default geometry described in the introduction b) \emph{``parallelepiped electrodes''}; c) \emph{``tilted electrodes''}; d) \emph{``trench electrodes''}
    }
    \label{fig:geometries}
\end{figure}

To test different geometries we have considered three configurations alternative to the geometry discussed in the previous 
sections and represented in Figure~\ref{fig:geometries}\textcolor{blue}{a}:
\begin{itemize}
    \item \emph{parallelepiped electrodes}, where all the electrodes are parallelepiped with square section and side 11~$\mu$m  (represented in Figure~\ref{fig:geometries}\textcolor{blue}{b}), this geometry essentially corresponds to the cylindrical configuration used in the \comsol field maps;
    \item \emph{tilted electrodes}, where the cylindrical polarization electrodes, with 12 $\mu$m diameter, are tilted 
        to favor field uniformity as proposed in Ref.~\cite{HuazhenLi:2024vci}
        (represented in Figure~\ref{fig:geometries}\textcolor{blue}{c});
    \item \emph{trench electrodes}, emulating the geometry of the TimeSPOT silicon sensors~\cite{Anderlini:2020ffm}. 
    This configuration is most probably unrealistic for 
        diamond without significant breakthrough in the graphitization, due to fabrication limitations  (trenching in 
        diamond is indeed prone to induce cracks in the material), but is reported here as an extreme limit to what 
        is achievable by the sole modification of the sensor geometry
        (represented in Figure~\ref{fig:geometries}\textcolor{blue}{d});

\end{itemize}

For each sensor geometry, including the standard one described in the previous sections, 
the same resistivity value (0.75~$\Omega \cdot \mathrm{cm}$) has been assigned to the 
graphitized regions, and the simulation of 1000 signals as generated by 180~GeV/$c$ pions 
impinging on the detector surface from randomized directions, are analyzed. 
To avoid possible biases in the interpretation of the comparisons between different geometries due to statistical fluctuations, the same set of pions, with the same ionization deposits,
and therefore the same distribution of carriers, have been used for all the simulations.

\begin{table}[h!]
\centering
\begin{tabular}{|c|c|c|}
\hline
 & \multicolumn{2}{|c|}{\textbf{Time resolution [ps]}} \\
\textbf{Sensor geometry} & \textbf{metallic electrodes} & \textbf{resistive electrodes} \\
\hline
Default with cylindrical electrodes ($\diameter = 12 \mathrm{\mu m}$) & $28 \pm 3$ & $71 \pm 4$ \\
\hline
Parallelepiped electrodes ($11 \times 11\ \mu\mathrm{m}^2$) & $28 \pm 1$ & $69 \pm 4$ \\
\hline
Tilted electrodes ($\diameter = 12\ \mu\mathrm{m}$) & $26 \pm 3$ & $69 \pm 4$ \\
\hline
Trench electrodes & $22 \pm 2$ & $50 \pm 3$ \\
\hline
\end{tabular}
\caption{Time resolution for the simulated geometries. The column ``metallic electrodes`` refers to a simulation assuming null resistivity of the electrodes, making the propagation of the signal (process 3) instantaneous. \methodC is used to compute the resolution in the column ``resistive electrodes'' where a resistivity $\rho = 0.75\, \Omega \cdot \mathrm{cm}$ is assumed for the electrodes.}
\label{tab:geometry-comparison}
\end{table}

The resolutions obtained from these simulation runs, assuming either perfectly conducting or graphitic electrodes, are reported in Table~\ref{tab:geometry-comparison}. The results with metallic electrodes provide an estimate of the contribution of field non-uniformities to the overall time resolution, whereas the results with a realistic resistivity represent a prediction of the sensor’s timing performance achievable with state-of-the-art graphitization technologies.

Comparing the results obtained with parallelepiped and cylindrical electrodes 
we notice that there is no significant difference, which implies that the exact shape 
of the electrode is not critical in the determination of the timing resolution. 

On the other hand, tilting the polarization electrodes proves beneficial for the uniformity of the electric field, without causing any degradation of the overall performance due to the increased resistance of the longer electrodes.

Metallic trench electrodes, extensively studied by the TimeSPOT collaboration~\cite{Anderlini:2020ffm}, are expected to provide the best performance in terms of field uniformity. This is confirmed by the simulation which estimates a time resolution close to the best theoretically achievable value given the assumed electronics and electronic noise. The residual difference is largely due to the loss of available charge carriers due to the thicker electrodes. Also, the larger thickness reduces the resistance of realistic graphitic electrodes improving the signal propagation characteristics as demonstrated by the significantly better time resolution. However, even with this extreme geometry, the electrode resistance remains the dominant contribution to the sensor timing performance.

\subsection{Time resolution and the effect of tilting angles}\label{sec:time-resolution}
The effects of the electrode resistance may extend beyond the simple increase of the jitter 
contribution.
For instance, the simulation predicts (and the beam test data confirm) that the asymptotic time 
resolution at very high signal amplitudes is still significantly worse than what is expected 
for metallic electrodes, despite the jitter contribution is drastically diminished because of the 
artificially large signal-to-noise ratio.

This additional degradation has been traced back to Landau fluctuations 
in the ionization energy deposits along the particle trajectory combined with a variable signal propagation delay that depends 
on the position at which the signal is induced along the readout electrode~\cite{Anderlini:2020ffm}.
Indeed, Landau fluctuations in the ionization deposits cause significant variability in the distribution of carrier densities along the electrode length, leading to a variable mixture of signal components subject to different delays. This variability in the charge distribution is then reflected in fluctuations of the rising edge of the signal and, ultimately, in the time stamp assigned to the waveform by the signal processing algorithm.

Therefore, even two identical particles traversing the sensor under the same conditions can produce signals 
with different leading-edge shapes. 
Higher electrode resistance amplifies these distortions, particularly at the signal leading edge, 
resulting in a spread of the timing measurements that is independent of the electronic noise.

The distribution of the charge induced along the electrode length is influenced by the tilt of the incoming particle relative to the readout electrodes. 
In silicon 3D sensors, where the timing resolution is mainly limited by inhomogeneities of the electrostatic field, tilting the sensor is generally beneficial. This is because it reduces the probability that the trajectory of an incoming particle lies entirely within either a high- or low-field region, thereby mitigating discrepancies between tracks, as most will traverse both types of regions.

In contrast, for 3D diamond sensors with resistive electrodes tilting the sensor causes particles to traverse high- and low-field regions at different depths, increasing the variability of propagation delays of induced signals along the electrode. Consequently, the leading edges of the signals become more variable, and the time resolution deteriorates as the inclination angle increases. 

The simulation developed in this work provides the opportunity to investigate the time resolution as a function of sensor inclination. In particular, it allows to disentangle two competing effects: one beneficial, related to the exploration of regions with different electric fields during the particle traversal, and one detrimental, arising from the tilt-induced variability of delayed signal components along the electrode.
This study is important for identifying the optimal operating angle that maximizes timing performance by balancing the competing effects of field inhomogeneities and fluctuations in charge deposition along the particle track.

\begin{table}[b!]
\centering
\begin{tabular}{|c|c|c|}
\hline
\textbf{$\theta$ [\textdegree]} & \textbf{Time resolution [ps]} \\
\hline
0 & $64.9 \pm 0.6$  \\
\hline
1 & $65.6 \pm 1.9$  \\
\hline
2 & $70 \pm 2$  \\
\hline
3 & $82 \pm 4$  \\
\hline
4 & $88 \pm 4$  \\
\hline
5 & $96 \pm 4$  \\
\hline
\end{tabular}
\caption{Time resolution as a function of the tilting angle of the sensor.}
\label{tab:tilted-angle}
\end{table}

\begin{figure}[tbp]
    \centering
    \begin{minipage}[c]{0.45\textwidth}
        \includegraphics[width=1.1\textwidth]{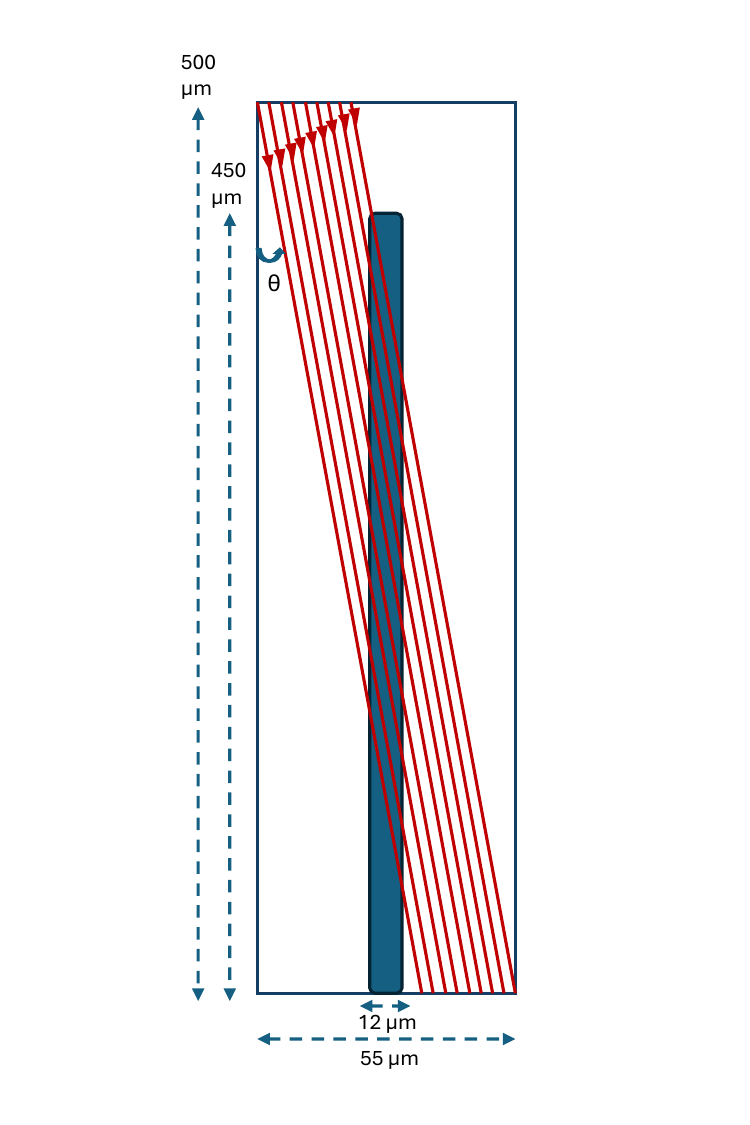}
        \caption{
            \label{fig:tilted-sensor}
            Schematic view of the method used to perform the study.
            }
    \end{minipage}
    \hfill
    \begin{minipage}[c]{0.51\textwidth}
        \includegraphics[width=\textwidth]{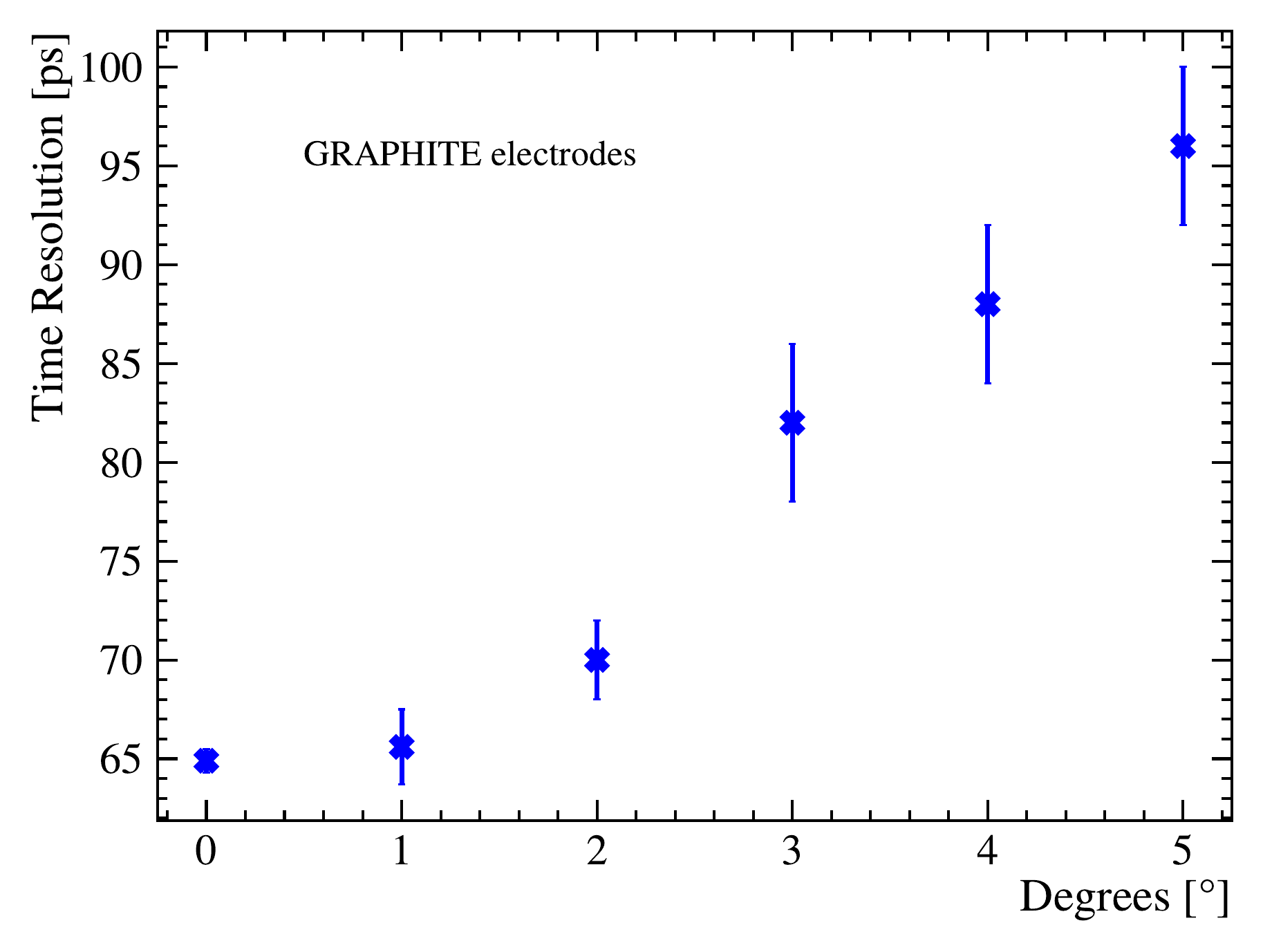}
        \caption{
            \label{fig:tilted-sensor-scan}
            Time resolution as a function of the sensor tilt angle for sensors with 
            resistive electrodes, using electric field maps obtained with \texttt{\comsol}.
            }
    \end{minipage}
\end{figure}

Since a single-pixel structure is simulated in this study, inclined tracks 
were generated ensuring that their trajectories remain entirely within 
the simulation volume. On the right side of Figure~\ref{fig:tilted-sensor}, a 
schematic illustration of this study for a generic inclination angle, $\theta$, 
is shown. As can be seen in 
Figure~\ref{fig:tilted-sensor-scan} and in Table~\ref{tab:tilted-angle}, the time resolution starts deteriorating already at very
small inclination angles, as the signal shape  becomes sensitive to
the depth profile of the charge deposition. 

This study highlights an effect of column resistance that extends 
beyond jitter and remains independent of electronic noise. 
Furthermore, it suggests that aligning diamond sensor electrodes with the 
expected trajectory of impinging particles may enhance timing resolution. 
In contrast, silicon 3D sensors exhibit improved performance when electrodes 
have an inclination of 10\textdegree\ or more with respect to the incident particles.
This configuration, however, promotes charge sharing between adjacent pixels, 
which can potentially degrade overall time resolution~\cite{Timespot:2023}.
 
\section{Conclusion}
\label{sec:conclusion}
In this work, we have provided a detailed analysis of the contributions to the timing 
uncertainty in 3D diamond detectors with graphitic electrodes.
To this end we have implemented a full simulation workflow exploiting a generalized version of the Ramo-Shockley theorem, developed for conductive media to 
account for signal propagation effects within the impedance network of a realistic
sensor.
Specifically, the simulation combines the modeling of ionization charge deposits performed with \texttt{Heed} and the computation of carrier trajectories implemented in \texttt{Garfield++},
with time-dependent weighting potentials obtained 
by numerically solving the \qsme either via FEMs using 
\texttt{\comsol} (\methodB) or through a custom solver leveraging spectral methods (\methodC).
The results obtained with these two alternative methods are reasonably consistent and qualitatively agree with expectations from a simplified 
model representing the sensor as a discrete impedance network (\methodA).

We validated the simulation against beam-test data from a 3D diamond sensor with electrodes of approximately 
30~k$\Omega$, corresponding to a resistivity of $0.75~\Omega \cdot \mathrm{cm}$.
The agreement between simulation and experimental data is acceptable, particularly considering the significant uncertainty in the electrode resistivity. 
Further validation and potential refinements of the simulation will be performed based on future measurements using geometries and experimental setups specifically designed for this purpose.

By varying the simulated electrode resistivity to assess the potential benefits of improved conductivity, we observed that 
already at 0.3~$\Omega \cdot \mathrm{cm}$ and below the contribution from the 
signal propagation becomes subdominant relative to the 
combined effects of jitter and field inhomogeneities.

We also explored the expected timing resolution for 
different geometries. 
Sensors with geometries beyond parallel columns -- such as those obtained
by tilting the electrodes -- reduce the contribution from electric field inhomogeneities; however, this effect remains negligible compared to the influence of signal propagation along the sensor depth and the expected electronic jitter.
A trench-structured sensor, with a geometry similar to the silicon 
sensors developed by TimeSPOT, would achieve a time resolution of approximately 50~ps, 
still dominated by the high resistivity of the electrodes.

Finally, we have used the simulation to assess the effect of resistive 
electrodes for particles impinging on the diamond sensor at angles 
between 0 and 5 degrees. The results indicate a rapid degradation of timing performance as the tilt increases.

This study highlights the importance of focusing future investigations on technological developments aimed at reducing electrode resistivity in the traditional parallel-column geometry.  
The results also indicate that a substantial improvement in the signal-to-noise ratio, along with optimized signal shaping, would greatly enhance the sensor's timing performance; thus, 
optimization of the readout electronics should proceed in 
parallel with studies of improved geometries. 

Future work will focus on automating the simulation flow and reducing systematic uncertainties related to mesh selection, while eagerly awaiting new experimental data for comparison.

\section*{Acknowledgments}

We are grateful to Andrea Lampis who, during the 2021 beam test, first proposed the idea that the timing resolution may worsen with the sensor tilt because 
of the resistance of the columns.

We are also grateful to Werner Riegler for his support and for his valuable suggestions.


\bibliographystyle{JHEP}
\bibliography{main.bib}




\end{document}